\documentclass[graybox]{svmult}
\usepackage{mathptmx}       
\usepackage{helvet}         
\usepackage{courier}        
\usepackage{type1cm}        
\usepackage{makeidx}         
\usepackage{graphicx}        
\usepackage{multicol}        
\usepackage[bottom]{footmisc}
\usepackage{amsmath}
\usepackage{amssymb}

\makeindex    

\newcommand{\si}{\sigma}
\newcommand{\al}{\alpha}
\newcommand{\tet}{\theta}
\newcommand{\az}{\varphi}
\newcommand{\ro}{\rho}

\newcommand{\be}{\beta}

\newcommand{\tro}{\tilde{\rho}}
\newcommand{\na}{\nabla}
\newcommand{\vsi}{\mbox{\boldmath{$\sigma$}}}
\newcommand{\vna}{\mbox{\boldmath{$\na$}}}
\newcommand{\Del}{\Delta}

\newcommand{\oeq}{\begin{equation}}
\newcommand{\ceq}{\end{equation}}
\newcommand{\oeqn}{\begin{eqnarray}}
\newcommand{\ceqn}{\end{eqnarray}}

\renewcommand{\>}{\rangle}
\newcommand{\<}{\langle}
\renewcommand{\(}{\left(}
\renewcommand{\)}{\right)}
\renewcommand{\[}{\left[}
\renewcommand{\]}{\right]}
\renewcommand{\lll}{\left|}
\newcommand{\rll}{\right|}

\newcommand{\stf}{\,\,\,}
\newcommand{\sdf}{\,\,}
\newcommand{\stb}{\!\!\!}

\newcommand{\kfi}{|\phi \>}

\newcommand{\ksp}{|s'\>}
\newcommand{\kvac}{|-\>}

\newcommand{\bfi}{\<\phi |}

\newcommand{\bs}{\< s |}

\newcommand{\oX}{\hat{X}}
\newcommand{\oY}{\hat{Y}}
\newcommand{\oP}{\hat{P}}

\newcommand{\oH}{\hat{H}}

\newcommand{\oro}{\hat{\rho}}
\newcommand{\oh}{\hat{h}}

\newcommand{\osi}{\hat{\sigma}}
\newcommand{\ovr}{\hat{\bf r}}

\newcommand{\ovR}{\hat{\bf R}}

\newcommand{\oad}{\hat{a}^\dagger}
\newcommand{\oa}{\hat{a}}

\newcommand{\oF}{\hat{F}}
\newcommand{\oN}{\hat{N}}

\newcommand{\ovsi}{\hat{\boldsymbol{\sigma}}}
\newcommand{\odel}{\hat{\delta}}
\newcommand{\ovk}{\hat{\bf k}}

\newcommand{\ov}{\hat{v}}

\newcommand{\del}{\delta\!}

\renewcommand{\d}{{\mbox d}}

\newcommand{\hb}{\hbar}

\newcommand{\vr}{{\bf r}}

\newcommand{\vj}{{\bf j}}
\newcommand{\vT}{{\bf T}}
\newcommand{\vJ}{{\bf J}}
\newcommand{\vk}{{\bf k}}

\newcommand{\vV}{{\bf V}}
\newcommand{\vW}{{\bf W}}
\newcommand{\vC}{{\bf C}}

\newcommand{\vS}{{\bf S}}

\newcommand{\ve}{{\bf e}}

\newcommand{\mH}{{\mathcal{H}}}

\newcommand{\mP}{{\mathcal{P}}}

\newcommand{\Tr}{\mbox{Tr}}
\newcommand{\tr}{\mbox{tr}}

\begin{document}
\title*{From light to hyper-heavy molecules and neutron-star crusts in a dynamical mean-field approach}
\titlerunning{Dynamical mean-field approach} 

\author{C\'edric Simenel}
\institute{CEA, Centre de Saclay, IRFU/Service de Physique Nucl\'eaire, F-91191 Gif-sur-Yvette, France, and\\  
 Department of Nuclear Physics, RSPE, Australian National 
University, Canberra, ACT 0200, Australia\\
\email{cedric.simenel@anu.edu.au}}

\maketitle

\abstract*{The richness of phenomena occurring in heavy-ion collisions calls for microscopic approaches where the motion of each nucleon is treated quantum mechanically. 
The most popular microscopic approach for low-energy collisions between atomic nuclei is the time-dependent Hartree-Fock (TDHF) theory, providing a quantum mean-field dynamics of the system. 
The TDHF approach and some of its extensions are used to predict the evolution of out-of-equilibrium nuclear systems. 
The formation of  di-nuclear systems with a structure close to molecular states is investigated. 
In particular, lifetimes and exit channels are described. 
The formation of light molecules and the dynamics of $\al$-clustering are discussed. 
Di-nuclear systems formed in transfer, deep-inelastic, and quasi-fission reactions, as well as hyper-heavy molecules produced in reactions between actinides are also investigated. 
The formation and stability of structures in neutron star crusts are finally discussed.
}

\abstract{The richness of phenomena occurring in heavy-ion collisions calls for microscopic approaches where the motion of each nucleon is treated quantum mechanically. 
The most popular microscopic approach for low-energy collisions between atomic nuclei is the time-dependent Hartree-Fock (TDHF) theory, providing a mean-field dynamics of the system. 
The TDHF approach and some of its extensions are used to predict the evolution of out-of-equilibrium nuclear systems. 
The formation of  di-nuclear systems with a structure close to molecular states is investigated. 
In particular, lifetimes and exit channels are described. 
The formation of light molecules and the dynamics of $\al$-clustering are discussed. 
Di-nuclear systems formed in transfer, deep-inelastic, and quasi-fission reactions, as well as hyper-heavy molecules produced in reactions between actinides are also investigated. 
The formation and stability of structures in neutron star crusts are finally discussed.}


\section{Introduction\label{sec:intro}}

Clustering in atomic nuclei is a general concept which includes a large variety of phenomena. 
Most of them have been covered in the volumes of ''Clusters in Nuclei''. 
These include, for instance, $\al$-clustering \cite{hor10,oer10,gup10,des12,yam12} 
and molecules formed by two light \cite{pap12,leb12,sal08}, 
intermediate/heavy \cite{ada12} 
or very heavy \cite{zag10,gup10} fragments. 

These cluster configurations are usually considered as specific structures of the total systems. 
However, except for light nuclei such as some beryllium isotopes \cite{hor10,kan10,gup10}, 
and eventually some heavier nuclei subject to cluster radioactivity \cite{poe10,gup10}, they are barely found in nuclear ground-state but rather in (sometimes highly) excited states. 
The question of the formation of these systems in nuclear reactions comes then naturally. 
However, the large variety of structures and reactions to investigate makes it very challenging to develop a unique model able to describe both dynamical and static properties of these systems. 

One possibility to describe clustering structures and dynamics is to assume {\it a priori} the presence of clusters in the state of the system. 
This is done, for instance, to describe $\al$-condenstates in Ref.~\cite{yam12} and di-nuclear systems in Ref.~\cite{ada12}. 
Combining microscopic and macroscopic degrees of freedom is also possible within the two-center shell model \cite{poe10,zag10} and the generator coordinate method (GCM) \cite{des12}. 

Nevertheless, purely microscopic approaches describing the state of each nucleon can be used assuming specific forms of the many-body wave functions. 
In the anti-symmetrised molecular dynamics (AMD) model, for instance, Gaussian single-particle wave-functions are considered \cite{hor10,kan10}. 
Another example of purely microscopic approach is the time-dependent Hartree-Fock (TDHF) theory (see Ref.~\cite{sim12b} for a recent review), which will be thoroughly used in the present chapter.


An attracting feature of the TDHF approach is that it uses the same energy density functional 
for both the structure of the collision partners and their dynamics during the reaction. 
Also, the same functional is used over the nuclear chart, allowing for 
both descriptions of structures and reaction mechanisms with a limiting number of parameters. 
In particular, only nuclear structure inputs are used in the fitting procedure of the functional. 

These aspects of the TDHF theory make it a promising tool to investigate various interplays between nuclear dynamics and (at least some) clustering effects. 
Few early TDHF codes have been used to study some cluster states and molecular structures. 
For instance, $\alpha$-clustering were investigated \cite{san83}, as well as light \cite{uma85} and hyper-heavy \cite{cus80,str83} molecules. 

More recently, the dynamics of di-nuclear states formed in heavy-ion collisions were analysed with modern three-dimensional TDHF codes. 
For instance, the path to fusion and nucleus-nucleus potentials have been studied in Refs.~\cite{kim97,sim01,sim04,mar06,uma06a,uma06b,uma06c,uma06d,guo07,sim07,uma07,sim08,guo08,uma08b,was08,uma09a,ayi09,uma09b,uma09c,was09a,uma10a,uma10b,obe10,loe11,iwa11,leb12,uma12a,obe12,kes12,uma12b,sim12b,sim12c,sim12d}. 
The transfer of one or many-nucleons and the isospin equilibration between the fragments in contact have also  been investigated in Refs. 
\cite{sim01,sim07,uma07,sim08,uma08a,was09b,iwa09,gol09,ked10,iwa10a,iwa10b,sim10b,eve11,yil11,sim11,iwa12,obe12,sim12a,sim12b,sim12c,sim12d}. 
In addition to heavy-ion collisions, the TDHF approach has been used to describe neutron star crust dynamics~\cite{seb09,seb11}.

First, formal and practical aspects of TDHF calculations are presented in Sec.~\ref{sec:TDHF}.
Then, the formation of light molecules and the dynamics of $\al$-cluster states are discussed in Secs.~\ref{sec:light_molec} and \ref{sec:al-cluster}, respectively. 
Quasi-elastic transfer is the subject of Sec.~\ref{sec:transfer}, followed by the study of more damped collisions, namely deep inelastic collisions in Sec.~\ref{sec:DIC} and quasi-fission reactions in Sec.~\ref{sec:QF}.
Calculations of hyper-heavy molecules dynamics  in reactions between actinides are presented in Sec.~\ref{sec:actinides}.
Finally, recent TDHF studies of neutron star crust dynamics are discussed in Sec.~\ref{sec:crust}.

\section{The time-dependent Hartree-Fock theory\label{sec:TDHF}}

The TDHF theory has been developed by Dirac in 1930~\cite{dir30}.
This is an extension of the mean-field approach to  the  ground-state of many-fermion systems introduced by Hartree~\cite{har28} and Fock~\cite{foc30}. 

\subsection{The mean-field approximation}

The TDHF theory determines the dynamics of a many-fermion system out of equilibrium under the approximation that the state of the system can be described by an independent-particle state at any time. 
The spatial correlations between the particles are obtained from the self-consistent mean-field. 
It is then assumed that each particle evolves independently in the mean-field generated by all the others. 

The TDHF approach is naturally well adapted to many-body systems in weak interactions. 
Indeed, when the interactions are strong, the system is expected to develop correlations which make the independent particle picture fail on a relatively short time scale.  
One may then wonder why the TDHF approach has been so successful in describing low-energy nuclear dynamics (see Refs.~\cite{neg82,sim12b} for reviews). 

In fact, at low energy, the Pauli principle prevents collisions between nucleons in such a way that the mean-free path of a nucleon in the nucleus is of the order of the size of the nucleus. 
This means that the wave-functions of the nucleons are essentially sensitive to the mean-field directly determined by the density. 
As a result, a nucleus in its ground state, where all the single-particle states below the Fermi level are almost entirely occupied, can be described with an independent particle state in a first approximation. 
Hartree-Fock calculations based on energy density functionals (EDF) are indeed able to reproduce quite well the binding energies and ground-state deformations along the nuclear chart (See Ref.~\cite{ben03} for a  review). 

Similarly, low-energy heavy-ion collisions can be treated at the mean-field level.
Indeed, at energies around the fusion barrier, the motion of the nuclei is slow enough to prevent nucleon-nucleon collisions thanks to the Pauli principle during the first few zeptoseconds (zs) of the reaction. 
However, the mean-field approximation is expected to fail at higher energies (e.g., in the Fermi regime), or for longer times. 
For instance, only the first steps of the fusion process can be described with TDHF, while beyond mean-field correlations are needed to form an equilibrated compound nucleus (CN) on a longer time scale. 

\subsection{Formalism}

The time-dependent Hartree-Fock equation reads~\cite{dir30}
\oeq
i\hb\frac{d}{dt}\rho(t) = \[ h[\rho(t)],\rho(t)\], 
\label{eq:TDHF}
\ceq
where $\rho(t)$ is the one-body density matrix with matrix elements 
\oeq
\rho_{\al\be} = \<\Phi| \oad_\be\oa_\al|\Phi\>.
\label{eq:rho}
\ceq
As the system is described by an independent particle-state, the state $|\Phi\>$ is a Slater determinant of the form
\oeq
|\Phi\> = \(\prod_{i=1}^A \oad_i\) \kvac,
\label{eq:Slater}
\ceq
where $A$ is the number of particles, $\oad_i$ creates an occupied single-particle state $|\varphi_i\>=\oad_i\kvac$, and $\kvac$ is the vacuum. 
The one-body density matrix associated to such a state is a projector onto the subspace of occupied single-particle states with the property $\rho^2=\rho$. It is associated to the single-particle operator
\oeq
\oro = \sum_{i=1}^A |\varphi_i\>\<\varphi_i|. 
\label{eq:rhoindep}
\ceq

The Hartree-Fock single-particle Hamiltonian $h[\rho]$ entering the TDHF equation~(\ref{eq:TDHF}) is obtained from the first derivative of the expectation value of the many-body Hamiltonian $\oH$ according to the one-body density matrix. 
Its matrix elements read
\oeq
h[\rho]_{\al\be} = \frac{\delta \<\Phi|\oH|\Phi\>}{\delta \rho_{\be\al}}.
\label{eq:hHF}
\ceq
In nuclear physics, however, the hard-core of the nucleon-nucleon interaction leads to a divergence of $\<\Phi|\oH|\Phi\>$ when evaluated on an independent-particle state. 
The effect of the hard-core can be renormalised in such a way that the energy does not diverge without affecting the low-energy properties of the system. 
In practical HF and TDHF calculations, the expectation value of $\oH$ is replaced by an energy density functional $E[\rho]$ determined, for instance, from the Skyrme \cite{sky56} or Gogny \cite{dec80} phenomenological effective interaction. 
The HF Hamiltonian then reads
\oeq
h[\rho]_{\al\be} = \frac{\delta E[\ro]}{\delta \rho_{\be\al}}.
\label{eq:hHFnuc}
\ceq

\subsection{The Skyrme energy density functional}

The Skyrme energy density functional is obtained from a zero-range effective interaction with gradient terms~\cite{sky56}. 
Numerical calculations are greatly helped by the zero-range nature of the interaction which simplifies the expression of the mean-field. 

The interaction between two nucleons in the medium reads
\oeqn
\ov (1,2) &=& t_0 \sdf \( 1+x_0\, \oP_\si \) \sdf \odel \nonumber \\
&+& \frac{1}{2} \sdf t_1 \sdf \( 1+x_1\, \oP_\si \) 
\sdf \(\ovk'^2 \sdf  \odel + \odel \sdf \ovk^2 \) \nonumber \\
&+& t_2 \sdf \( 1+x_2\, \oP_\si \) 
\sdf \(\ovk' \cdot  \odel \sdf  \ovk \) \nonumber \\
 &+& \frac{1}{6} \sdf t_3 \sdf \( 1+x_3\, \oP_\si \) 
 \sdf \ro^\al\! (\ovR) \sdf \odel \nonumber \\
 &+& i\, W_0 \sdf \ovsi \cdot \(\ovk' \times \odel \, \ovk \) 
\label{eq:skyrme}
\ceqn
where $\odel = \del\(\ovr(1)-\ovr(2)\)$, $\ovk = \frac{1}{2i}\(\vna(1)-\vna(2)\)$ 
(relative momentum), $\ovk'$ is the complex conjugated of $\ovk$ acting on the left, and  $\ovR = \(\ovr(1)+\ovr(2)\)/2$.
The operators $\ovsi = \ovsi(1)+ \ovsi(2)$, with
$\ovsi(i) = \osi_x\!(i) \, \ve_x+ \osi_y\!(i) \, \ve_y + \osi_z\!(i) \, \ve_z$, 
are expressed in terms of the Pauli matrices $\osi_{x/y/z}(i)$ 
acting on the spin of the particle $i$.
 $\oP_\si = \[1+  \ovsi(1) \cdot \ovsi(2) \]/2$ corresponds to the exchange 
of the spins.  The particle density in~$\vr$
is noted $\ro(\vr) \equiv \sum_{sq} \ro(\vr s q ,\vr s q )$ where $\rho$ is the one-body density matrix, $s$ the spin and $q$ the isospin.
The "$t_1$" and "$t_2$" terms are non-local in space and simulate the short range part of the interaction.
Finally the last term accounts for the spin-orbit interaction.

The EDF describes how the energy of the system depends on its density. 
It is standard to introduce a local energy density $\mH(\vr)$ such that
\oeq
E=\int d\vr \mH(\vr).
\ceq
This energy density can be expressed as~\cite{bon87}
\oeqn
\mH(\vr) &=& \frac{\hb^2}{2m}\tau + B_1 \rho^2 + B_2 \sum_q\rho_q^2 \nonumber \\
&&+B_3(\rho \tau -\vj^2)+B_4\sum_q(\rho_q \tau_q -\vj_q^2)\nonumber \\
&&+B_5\rho\Delta\rho+B_6\sum_q\rho_q\Delta\rho_q+B_7\rho^{2+\alpha}+B_8\rho^\al \sum_q\rho_q^2\nonumber\\
&&+B_9(\rho\vna\!\cdot\!\vJ + \vj\!\cdot\! \vna\!\times \!\vS+\sum_q\rho_q\vna\!\cdot\!\vJ_q+\vj_q\!\cdot\!\vna\!\times\!\vS_q)\nonumber\\
&&+B_{10} \vS^2+B_{11}\sum_q\vS_q^2 +B_{12}\rho^\al \vS^2 +B_{13} \rho^\al \sum_q\vS_q^2.\nonumber \\
&&\label{eq:Hr}
\ceqn

The densities entering Eq.~(\ref{eq:Hr}) are the local density
\oeq
\ro(\vr) = \sum_{i s}\sdf  \az_{i}^*(\vr s)\,  \az_{i}(\vr s) ,   
\ceq 
where $s$ denotes the spin, and $\az_i$ are occupied single-particle states, the kinetic energy density
\oeq
\tau(\vr) =   \sum_{i s}\sdf  |\vna \az_i(\vr s)|^2,  
\ceq
the current density
\oeq
\vj(\vr) = \frac{1}{2\,i} \sum_{i s} \az_i^*(\vr s) \sdf \vna \sdf \az_i(\vr s) \stf +c.c., 
\ceq
where $c.c.$ means "complex conjugated", the gradient of the spin-orbit density
\oeq
\vna . \vJ(\vr) = -i\sum_{i ss'}   \vna\az_i^*(\vr s) \times \vna \az_i (\vr s')\cdot \bs \vsi \ksp ,
\ceq
and the spin density
\oeq
\vS(\vr) = \sum_{i s} \sdf  \az_i^*(\vr s) \sdf \az_i (\vr s')\sdf \bs \vsi \ksp.
\ceq
In the above expressions, it is assumed that the one-body density-matrix $\ro$ is diagonal in isospin.
The isospin is then omitted to simplify the notation.
The $\vj$ and $\vS$ densities are time-odd and vanish in time-reversal invariant systems. 
They are, however, important in time-dependent calculations to ensure Galilean invariance~\cite{eng75}. 

Note that the general Skyrme EDF includes other terms which are neglected in Eq.~(\ref{eq:Hr}). 
These additional terms are of the form $\vS\cdot\Delta\vS$ and with other densities, i.e., the spin-current pseudo-tensor $\stackrel{\leftrightarrow}{J}$ and the spin-kinetic energy density $\vT$~\cite{eng75,uma06a}. They are sometimes included in TDHF calculations~\cite{uma06a,mar06,loe12}. In Eq.~(\ref{eq:Hr}), only the anti-symmetric part of $\stackrel{\leftrightarrow}{J}$, which is the spin-orbit density $\vJ$, is  included. The spin-orbit energy is indeed expected to be more important (by about one order of magnitude) than the other spin-gradient terms~\cite{cha98}.

The coefficients $B_i$ in Eq.~(\ref{eq:Hr}) are related to the parameters of the Skyrme effective interactions $\{t_{0-3},x_{0-3},\al,W_0\}$ as
\oeqn
B_1&=&\frac{t_0}{2}\(1+\frac{x_0}{2}\)\nonumber\\
B_2&=&-\frac{t_0}{2}\(x_0+\frac{1}{2}\)\nonumber\\
B_3&=&\frac{1}{4}\[t_1\(1+\frac{x_1}{2}\)+t_2\(1+\frac{x_2}{2}\)\]\nonumber\\
B_4&=&-\frac{1}{4}\[t_1\(x_1+\frac{1}{2}\)-t_2\(x_2+\frac{1}{2}\)\]\nonumber\\
B_5&=&-\frac{1}{16}\[3t_1\(1+\frac{x_1}{2}\)-t_2\(1+\frac{x_2}{2}\)\]\nonumber\\
B_6&=&\frac{1}{16}\[3t_1\(x_1+\frac{1}{2}\)+t_2\(x_2+\frac{1}{2}\)\]\nonumber\\
B_7&=&\frac{t_3}{12}\(1+\frac{x_3}{2}\)\nonumber\\
B_8&=&-\frac{t_3}{12}\(x_3+\frac{1}{2}\)\nonumber\\
B_9&=&-\frac{1}{2}W_0\nonumber\\
B_{10}&=&\frac{t_0x_0}{4}\nonumber\\
B_{11}&=&-\frac{t_0}{4}\nonumber\\
B_{12}&=&\frac{t_3x_3}{24}\nonumber\\
B_{13}&=&-\frac{t_3}{24}.\nonumber\\
\ceqn

The Skyrme-HF mean-field is derived from Eq.~(\ref{eq:hHFnuc}). 
Its action on single-particle wave functions is then given by \cite{bon87}
\oeqn
&&\(h[\ro]  \az_i \)(\vr, s ) = \nonumber\\
&&\sum_{s'} \!\[\! \(\!-\vna \frac{\hb^2}{2m^*_{q_i}\!(\vr)} \vna \!+\! U_{q_i}\!(\vr)\! +\! i\vC_{q_i}\!(\vr)\! \cdot\! \vna\! \)\!\delta_{ss'}\right. \nonumber \\
&&\left. + \vV_{q_i}\!(\vr)\cdot  \bs \vsi \ksp 
 +i\vW_{q_i}\!(\vr) \cdot \( \bs \vsi \ksp \times \vna \)\frac{}{} \] \az_i(\vr,s'),
 \nonumber \\
\label{eq:HFskyrme}
\ceqn
where $q_i$ is the isospin of the state $|\az_i\>$.
The derivatives act on each term sitting on their right, including the wave function.
The fields (functions of $\vr$) read
\oeqn
\frac{\hb^2}{2\,m_q^*} &=& \frac{\hb^2}{2\,m} + B_3 \,\ro + B_4\, \ro_q \label{eq:effmass}\\
U_q &=& 2B_1\ro+2B_2\ro_q+B_3(\tau+i\vna\cdot\vj)+B_4(\tau_q+i\vna\cdot\vj_q)\nonumber\\
&&+2B_5\Delta\ro+2B_6\Delta\ro_q+(2+\al)B_7\ro^{1+\al}\nonumber\\
&&+B_8[\al\ro^{\al-1}\sum_q\ro_q^2+2\ro^\al\ro_q]+B_9(\vna\cdot \vJ +\vna \cdot \vJ_q)\nonumber \\
&&+\al\ro^{\al-1}(B_{12}\vS^2+B_{13}\sum_q\vS_q^2)\\
\vV_q&=&B_9\vna\times (\vj+\vj_q)+2B_{10}\vS+2B_{11} \vS_q\nonumber\\
&&+2\rho^\al(B_{12}\vS+B_{13}\vS_q)\\
\vW_q &=& -B_9 \, \vna \, \(\ro+ \ro_q \)\label{eq:W}\\
\vC_q &=& 2 \, B_3 \, \vj + 2 \, B_4 \, \vj_q - B_9 \,  \vna \times \(\vS + \vS_q \) ,\label{eq:C} 
\ceqn
where the derivatives act on the first term sitting on their right only. 
The label $q$ denotes the isospin. 
The effective mass $m_q^*$ of nucleons with isospin $q$ is introduced in [Eq.~(\ref{eq:effmass})]. 
It originates from the non-local terms of the effective interaction in Eq.~(\ref{eq:skyrme}).

The parameters of the Skyrme EDF are fitted on few quantities (see, e.g., Ref.~\cite{cha98}). 
These include the density $\rho_0\simeq0.16$~fm$^{-3}$ and energy  per nucleon $E/A\simeq-16$~MeV of the infinite symmetric nuclear matter at saturation as well as its compressibility. 
Depending on the parametrisation, the equation of state of the infinite neutron matter \cite{wir88}, the enhancement factor of the Thomas-Reiche-Kuhn sum rule, the symmetry energy, and the radii and binding energies of few doubly-magic nuclei may be included as constraints into the fitting procedure as well. 

It is interesting to note that no input on nuclear reaction mechanism, such as fusion barriers or cross-sections, are included in the fit. 
Nevertheless, as we will see in this chapter, the description of collision dynamics with TDHF is very realistic, and agreements with experimental observables are sometimes impressive. 

Finally, the Coulomb interaction between the protons is added to the Skyrme mean-field.
The direct part of the Coulomb energy reads
\oeq
E_c^{dir}=\frac{ e^2}{2} \int d^3r \int d^3r' \frac{\rho_p(\vr)\rho_p(\vr')}{|\vr-\vr'|}.
\ceq
The latter is usually computed by solving, first, the Poisson equation to get the Coulomb potential $V_c(\vr)$, and, then, by evaluating the integral $\frac{1}{2}\int d^3r \rho_pV_c$.
The exchange part of the Coulomb energy is usually determined within the Slater approximation as
\oeq
E_c^{ex}=\frac{-3e^2}{4}\(\frac{3}{\pi}\)^{\frac{1}{3}} \int d^3r \rho_p(\vr)^\frac{4}{3}.
\ceq
As a result, the contribution of the Coulomb interaction to the proton mean-field reads 
\oeq
U_c=V_c-e^2\(\frac{3\rho_p}{\pi}\)^{\frac{1}{3}}.
\ceq

\subsection{Numerical implementation}

The TDHF equation is never implemented with its Liouville-von Neumann form given in Eq.~(\ref{eq:TDHF}). 
Instead, a set of non-linear Schr\"odinger-like equations for single-particle motion is used. 
Indeed, Eq.~(\ref{eq:TDHF}) can be expressed in a fully equivalent way as
\oeq
i\hb \frac{d}{dt} |\az_i(t)\> = \oh[\ro(t)] |\az_i(t)\>, \stf 1\le i\le A .\label{eq:TDHFsp}
\ceq

The numerical advantage of using Eqs.~(\ref{eq:TDHFsp}) instead of Eq.~(\ref{eq:TDHF}) is obvious in terms of computer memory. 
Indeed, for a basis of $N$ single-particle states (e.g., the number of points of a cartesian grid), the storage of $\rho$ requires a $N\times{N}$ array, while the wave-functions in Eq.~(\ref{eq:TDHFsp}) require a $N\times{A}$ array. 
As $A$ is usually much smaller than $N$, one clearly realises the advantage of solving Eqs.~(\ref{eq:TDHFsp}) instead of Eq.~(\ref{eq:TDHF}).

Equations~(\ref{eq:TDHFsp}) are coupled by the self-consistency of the HF Hamiltonian as it depends on the total density of the system. 
As a result, the HF Hamiltonian is time-dependent and one needs to solve Eqs.~(\ref{eq:TDHFsp}) iteratively in time. 
The sates at time $t+\Delta t$ are determined from the states at time $t$ assuming that $\oh$ is constant between $t$ and $t+\Delta t$. This implies that $\Delta t$ has to be chosen small enough for this condition to be valid. 
Typical time step increments $\Delta t$ in nuclear physics range from $\sim5\times10^{-25}$~s \cite{nak05,sim09} to $\sim1.5\times10^{-24}$~s \cite{kim97,uma05}.

In addition, to conserve energy and particle number, the algorithm has to be symmetric under time-reversal transformation.
This implies that the HF Hamiltonian has to be evaluated at $t+\frac{\Delta{t}}{2}$~\cite{bon76}. 
The evolution operator then reads
\oeq
|\az_i(t+\Delta t)\> \simeq e^{i\oh(t+\frac{\Delta t}{2})/\hb} |\az_i(t)\> \label{eq:evolop}.
\ceq
A truncated Taylor development of the exponential is usually considered. 
The evolution operator then breaks unitarity and one should check the orthonormalisation of the wave-functions during the time evolution.  

A possible algorithm to perform the time evolution in Eq.~(\ref{eq:evolop}) is described below:
\oeq
\begin{array}{ccc}
\{ |\az_1^{(n)}\> \cdots |\az_A^{(n)} \>\} & \Rightarrow &\ro^{(n)}\\
\Uparrow && \Downarrow\\
|\az_i^{(n+1)}\> =  e^{-i\frac{\Delta t}{\hb}  \oh^{(n+\frac{1}{2})}}  |\az_i^{(n)}\>
 && \oh^{(n)}\equiv\oh[\ro^{(n)}] \\
\Uparrow && \Downarrow\\
 \oh^{\(n+\frac{1}{2}\)}\equiv\oh\[\ro^{\(n+\frac{1}{2}\)}\] &&
   |\tilde{\az}_i^{(n+1)}\> = e^{-i\frac{\Delta t}{\hb} \oh^{(n)}}  |\az_i^{(n)}\> \\ 
\Uparrow & &   \Downarrow \\
\ro^{\(n+\frac{1}{2}\)}= \frac{\ro^{(n)} + \tro^{(n+1)}}{2} & \Leftarrow & \tro^{(n+1)} \\
\end{array} 
\label{eq:algo}
\ceq
where $|\az_i^{(n)}\>$ is an approximation of $|\az_i(t_n=n\Del t)\>$.
A first evolution over $\Delta{t}$ is performed to estimate the density at $t+\Delta{t}$.
The latter is used, together with the density at $t$, to determine the density, and then the HF Hamiltonian, at $t+\frac{\Delta{t}}{2}$. 
This Hamiltonian is finally used to evolve the wave-function from $t$ to $t+\Delta{t}$. 

Possible single-particle bases  to solve the TDHF equation numerically are the harmonic-oscillator basis~\cite{has12}, basis-spline collocation method~\cite{uma91}, wavelets~\cite{seb09}, adaptive networks~\cite{nak05}, or regular cartesian grids~\cite{kim97,mar05}. 
Typical regular mesh spacing with $\Delta x\simeq0.6$~fm \cite{sim09}, 0.8~fm \cite{kim97} and 1.0~fm \cite{uma05,mar05} are used.

The initial condition of a TDHF calculation of heavy-ion collisions usually assumes that the nuclei are at some finite distance in their HF ground state. 
HF calculations of the collision partners then need  to be performed prior to the TDHF evolution. 
This is done with the same EDF as in the TDHF calculation to ensure full self-consistency between structure and dynamics.
Large initial distances between the centers-of-mass should be used to enable a proper treatment of the Coulomb excitation in the entrance channel. Typical distances of the order of $\sim40$~fm are considered as a good compromise to limit computational time.  
It is also usually assumed that the nuclei followed a Rutherford trajectory prior to this initial condition. 
It determines the initial momenta $\hb\vk$ to be applied to the nucleons using Galilean boosts of the form
\oeq
|\az_i(t=0)\> = e^{-i\vk\cdot\ovr}|\az_i^{HF}\>,
\ceq
where $|\az_i^{HF}\>$ are the HF single particle states. 

More details on numerical implementations of the TDHF equation can be found in Refs.~\cite{sim10a,sim12b}.

\subsection{Beyond the TDHF approach}

The independent-particle approximation can be considered as a zeroth order approximation to the many-body problem. 
In fact,  the exact evolution of the one-body density-matrix reads
\oeq
i\hbar \frac{\partial }{\partial t}\rho _{1}
=\left[ t_1,\rho _{1}\right] + \frac{1}{2}{\rm \Tr}_{2}\left[ \bar v_{12},\rho_{12}\right]  ,
\label{eq:TDDM}
\ceq
where $\bar{v}_{12}$ is the antisymmetrised two-body interaction and $\rho_1$ and $\rho_{12}$  are the one- and two-body density-matrices, respectively.
Solving Eq.~(\ref{eq:TDDM}) requires the knowledge of $\rho_{12}(t)$. 
The latter obeys an evolution equation which depends on the three-body density-matrix. 
In fact, Eq.~(\ref{eq:TDDM}) is the  first equation of the BBGKY hierarchy~\cite{bog46,bor46,kir46} providing a set of coupled equations for $\rho_{1}$, $\rho_{12}$, $\rho_{123}$...

We see that the TDHF equation is obtained by neglecting the last term in Eq.~(\ref{eq:TDDM}). 
It is important to know what is the physical meaning of this term. 
It contains the so-called two-body correlations which develop because of the residual interaction, i.e., the difference between the exact and mean-field Hamiltonians. 
Three main types of correlations can be identified:
\begin{itemize}
\item pairing correlations,
\item correlations induced by a collision term,
\item and long-range dynamical fluctuations.
\end{itemize}

Pairing correlations are important for a proper description of mid-shell nuclei, as well as to describe pair-transfer reactions, as we will see in Sec.~\ref{sec:transOPb}. 
They can be included using a ''generalised'' mean-field approximation. 
In this case, the state of the system is described as a quasi-particle vacuum~\cite{rin80}.
This leads to the BCS model for pairing between time-reversed states, or, more generally, to the Hartree-Fock-Bogoliubov approximation. 
Nuclear dynamics in presence of pairing has been investigated recently with the TD-BCS approach \cite{eba10,sca12}, and at the TDHFB level \cite{ave08,ste11,has12}. 
The linearised version of TDHFB is the quasi-particle random-phase approximation (QRPA) which has been widely applied to study nuclear vibrations \cite{eng99,kha02,fra05,per08}. 
Applications of the TDHFB formalism to study pairing vibrations are presented in Sec.~\ref{sec:pairvib}. 

The collision term is important at high energy, and to describe long-term dynamics such as the thermalisation of the compound nucleus. 
It can be added to the TDHF equation in what becomes the Extended-TDHF formalism \cite{won78,won79,dan84,bot90,ayi80,lac99}. 
A more general approach including pairing and a collision term is given by the time-dependent density-matrix (TDDM) formalism \cite{cas90,bla92} which has been applied to heavy-ion collisions~\cite{toh02b,ass09}.
However, to describe reaction mechanisms at intermediate  energy such as multi-fragmentation, quantum effect can be neglected in a first approximation. 
Semi-classical versions of the mean-field theory including collision terms, such as the Landau-Vlasov formalism, have then been widely used to describe reaction mechanisms at intermediate energy \cite{gre87,sch89,mot92}.

Long-range dynamical fluctuations may play an important role even at low energy. 
For instance, they are crucial to determine fluctuations of one-body operators, such as the fragment mass and charge distribution widths in heavy-ion collisions. 
In the limit where fluctuations around the TDHF path are small, then a good description of these distributions is obtained within the time-dependent RPA (TDRPA) formalism. 
The latter can be obtained from the Balian-V\'en\'eroni variational principle \cite{bal81,bal84}. 
Numerical applications to describe fragment mass and charge distributions in deep-inelastic collisions have been recently performed and compared to experiment in Ref.~\cite{sim11}. 
Alternatively, such fluctuations could also be obtained from the stochastic mean-field (SMF) approach~\cite{ayi08,was09b}, or from the time-dependent generator coordinate method (TDGCM)~\cite{rei83,gou05}. 

Realistic calculations with these approaches beyond TDHF are often very demanding from a numerical point of view and systematic applications are usually prohibitive even with modern high-performance computing facilities. 
In this chapter, we then focus on TDHF applications, paying attention to the limitations and possible improvements in the future.  
However, recent numerical calculations with the TDHFB and TDRPA approaches are also presented.

\section{Formation of light molecules \label{sec:light_molec}}

Nuclear molecules made of light nuclei such as $^{12}$C and $^{16}$O are formed and studied thanks to nuclear  collisions.
The dynamics of the formation of such di-nuclear systems is the purpose of the present section.

\subsection{Structures in fusion cross-sections}

Structures in fusion cross-sections are possible experimental signatures of nuclear molecules \cite{sch83,leb12}.
However, structures in fusion excitation functions may also appear in light systems which are not necessarily due to the formation of nuclear molecules.
Such structures or oscillations appear clearly in cross-sections for the fusion of $^{12}$C+$^{12}$C \cite{spe76b}, $^{12}$C+$^{16}$O \cite{spe76a}, and $^{16}$O+$^{16}$O \cite{kol77,tse78,kov79}. 
In particular, the discrete nature of angular momentum may reveal itself in fusion excitation functions as peaks associated to barriers for specific angular momenta \cite{van79,esb08,esb12}. 

The fusion cross-section is written as
\oeq
\si_{fus.}(E_{c.m.}) = \frac{\pi\hb^2}{2\mu E_{c.m.}} \sum_{L=0}^\infty (2L+1) P_{fus.}(L,E_{c.m.})
\ceq
where $\mu$ is the reduced mass of the system. $P_{fus.}(L,E_{c.m.})$ is the fusion probability for the partial wave with orbital angular momentum $L$ at the center-of-mass energy $E_{c.m.}$. 

TDHF calculations do not include tunnelling of the many-body wave-function, i.e., $P_{fus.}^{TDHF}=0$ or~1. 
As a result, the fusion cross-section can be estimated with the quantum sharp cut-off formula \cite{bla54}
\oeqn
\si_{fus.}(E_{c.m.}) &=& \frac{\pi\hb^2}{2\mu E_{c.m.}} \sum_{L=0}^{L_{max}(E_{c.m.})} (2L+1) \nonumber \\
&=&  \frac{\pi\hb^2}{2\mu E_{c.m.}} \[L_{max}(E_{c.m.})+1\]^2,
\ceqn
where $L_{max}(E_{c.m.})$ is the maximum angular momentum at which fusion occurs at $E_{c.m.}$. 
For fusion of symmetric systems with $0^+$ ground-states, fusion can only occur for even values of the angular momentum. 
The cross-section with the sharp cut-off formula then reads
\oeq
\si_{fus.}(E_{c.m.}) = \frac{\pi\hb^2}{2\mu E_{c.m.}} \[L_{max}(L_{max}+3)+2\].
\label{eq:cseven}
\ceq

\begin{figure}
\sidecaption
\includegraphics[width=7.5cm]{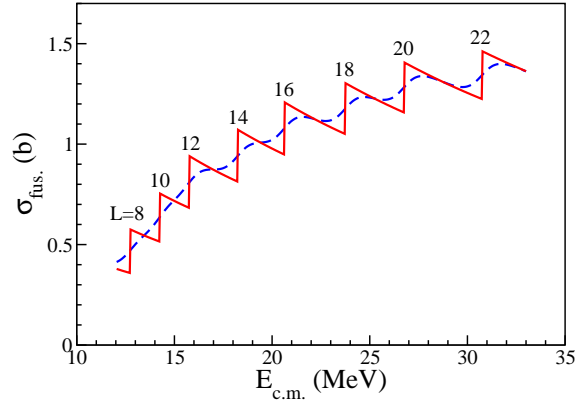}
\caption{Fusion cross-section as a function of center-of-mass energy in $^{16}$O+$^{16}$O obtained with TDHF calculations. The cross-sections are computed with the sharp cut-off formula (solid line) and using Eq.~(\ref{eq:PfusHW}) for the barrier penetration probabilities (dashed line). The numbers indicate the position of the barriers $B(L)$.
\label{fig:csOO}}
\end{figure}

An example of fusion cross-section in $^{16}$O+$^{16}$O as function of energy obtained with the \textsc{tdhf3d} code with the SLy4$d$ Skyrme EDF \cite{kim97} and Eq.~(\ref{eq:cseven}) is shown in Fig.~\ref{fig:csOO} with solid line. 
The sharp increases of the fusion cross-sections at the positions of the angular momentum dependent barriers $B(L)$ are due to the fact that fusion penetration probabilities are either 0 or 1 at the TDHF level. 

These peaks are highly smoothen when tunnelling is taken into account. 	
As a first approximation, one can estimate the barrier penetration probability according to the Hill-Wheeler formula \cite{hil53} with a Fermi function 
\oeq
P_{fus.}(L,E_{c.m.})\simeq \frac{e^{x_L}}{1+e^{x_L}},
\label{eq:PfusHW}
\ceq
with $x_L=[E-B(L)]/\varepsilon$. 
Choosing the decay constant $\varepsilon=0.4$ \cite{esb12}, one gets the fusion cross-sections represented by a dashed line in Fig.~\ref{fig:csOO}. 
Oscillations for $E_{c.m.}>16$~MeV are clearly visible and due to $L$-dependent barriers with $L\ge12\hb$.  
Note that these oscillations are less visible for asymmetric systems due to the fact that all integer values of $L$ are possible. 
In addition, the observation of these oscillations is limited to light systems up to, e.g., $^{28}$Si+$^{28}$Si \cite{gar82,esb12}. 
For heavier systems, the oscillations are indeed expected to be smeared out as the coupling to many reaction channels sets in \cite{esb12}. 

To conclude, structures due to oscillations of the fusion cross-sections generated by the discrete nature of the angular momentum are expected to occur in light systems, in particular for symmetric collisions. 
As a result, one should be careful in the search for resonances associated to molecular states in these systems.
In particular, the  observation of a peak in the fusion cross-sections may not be sufficient to assign such a structure to a resonance state. One should, in addition, search for this resonance in other channels, and investigate its decay properties \cite{leb12}. 

\subsection{Contact times around the barrier in $^{12}$C+$^{16}$O}

Resonances at and below the barrier may strongly affect the reaction outcome.
For instance, narrow resonances have been observed in the radiative capture $^{12}$C($^{16}$O,$\gamma$)$^{28}$Si close to the Coulomb barrier~\cite{leb12}.  
In particular, tentative spins of $4\hb$, $5\hb$ and $6\hb$ have been assigned in the collision energy range $E_{c.m.}\simeq8.5-9$~MeV \cite{leb12,sch83,tre78}.

To try to better understand in a dynamical way the presence of resonances and the contribution of relatively large spins at energies close to the Coulomb barrier,  TDHF calculations have been performed on the $^{12}$C+$^{16}$O system \cite{leb12}.
For this system, the Coulomb barrier obtained with the \textsc{tdhf3d} code and the SLy4$d$ Skyrme functional \cite{kim97} is  $B=7.85\pm0.05$~MeV.

\begin{figure}
\sidecaption
\includegraphics[width=7.5cm]{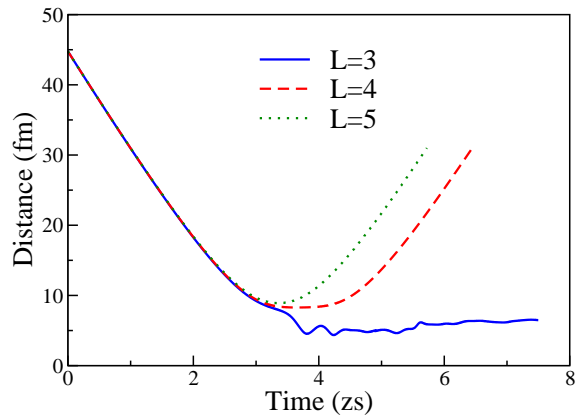}
\caption{Distance between the centers-of-mass of the fragments as a function of time in $^{12}$C+$^{16}$O at $E_{c.m.}=8.8$~MeV for $L=3\hb$, $4\hb$ and $5\hb$. 
\label{fig:distCO}}
\end{figure}

Fig.~\ref{fig:distCO} shows the distance between the centers-of-mass of the fragments as a function of time at $E_{c.m.}=8.8$~MeV for different values of the angular momentum. 
Fusion occurs at $L\le3\hb$ and re-separation at $L\ge4\hb$. 
Without the presence of molecular states at $4\hb$, $5\hb$ and 6$\hb$, the system would undergo a fast re-separation within $\sim1-2$~zs. 
This indicates that these spins are populated by a direct transition toward resonant states of the compound nucleus. 

In the collision, the system may spend some time in a di-nuclear configuration, which presents some analogies with a molecular state. 
The excitation of the latter is then expected to increase with the lifetime of the di-nuclear system. 
The definition chosen here for the existence of a di-nucleus is the following: the nuclear density at the neck should be between 0.004 and 0.14~fm$^{-3}$, i.e., lower than the saturation density of 0.16~fm$^{-3}$. 
These nuclear densities correspond to distances between $^{12}$C and $^{16}$O from 5.98 to 10.43~fm. 

\vspace{0.5cm}

\begin{figure}
\sidecaption
\includegraphics[width=7.5cm]{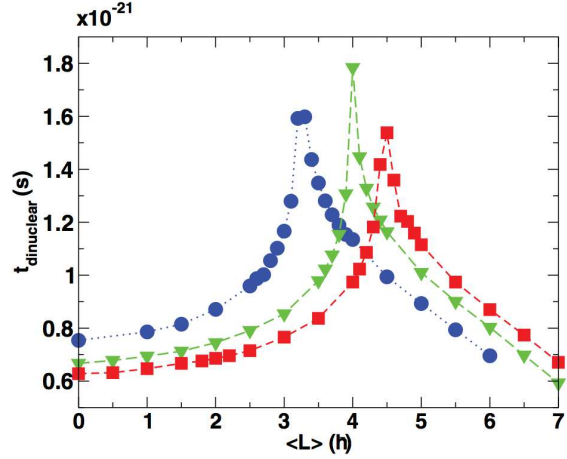}
\caption{Di-nuclear lifetimes (see text) as a function of the mean angular momentum at $E_{c.m.}=8.5$~MeV  (blue circles), 8.8~MeV (green triangles), and 9~MeV (red squares). Adapted from Ref.~\cite{leb12}.}
\label{fig:contact_CO}
\end{figure}

The di-nuclear lifetime is shown as a function of the average angular momentum $\<L\>$ in Fig.~\ref{fig:contact_CO}. 
Peaks are observed at the critical mean angular momentum  $\<L\>_c$ for fusion. 
We get $\<L\>_c\simeq3.2\hb$, $3.9\hb$, and $4.4\hb$ for $E_{c.m.}=8.5$, 8.8, and 9~MeV, respectively. 
Below $\<L\>_c$, this time increases with $\<L\>$ because the fusion process is slowed down by the centrifugal repulsion. 
In contrast, for $\<L\>$ greater than $\<L\>_c$, the fragments re-separate and the time of contact decreases for more peripheral collisions. 

With the present definition of the di-nuclear system lifetime, we see that such a system exists at relatively high angular momenta of $5\hb$ and $6\hb$ during $\sim0.7-1.2$~zs depending on the energy. 
The corresponding lifetimes are shorter than the lifetimes of typical resonances at the Coulomb barrier, which are $\sim2.6$~zs \cite{erb85,tre78}. 
However, they may be sufficient to enable a direct excitation of a resonant or molecular state of the compound nucleus with a similar structure. 
In particular, we see that, for $\<L\>=5\hb$ and $6\hb$ (above the critical angular momenta), the lifetime increases with energy. 
Similarly, the probability for the population of a molecular state is also expected to increase. 
This may explain why the experimental cross-section for radiative capture is observed to increase with energy for these spins~\cite{leb12}.

\subsection{The $J^\pi=36^+$ resonance in $^{24}$Mg+$^{24}$Mg}

The $^{24}$Mg+$^{24}$Mg system presents a resonance with high spin ($36\hb$ to $38\hb$) at twice the Coulomb barrier, corresponding to an excitation energy $E^*\simeq60$~MeV in the $^{48}$Cr. 
At this energy, many decay channels are open and it is necessary, from an experimental point of view, to investigate as many of these channels as possible. 
For instance, the $J^{\pi}=36^+$ resonance in $^{24}$Mg+$^{24}$Mg is observed at $E_{c.m.}=45.7$~MeV essentially in inelastic scattering \cite{zur83} with a lifetime $\tau=\frac{\hb}{\Gamma}\simeq3.9$~zs.  
The decay of this resonance has been recently investigated in both inelastic and fusion-evaporation channels \cite{sal08}. 

One particularity of the $^{24}$Mg is its strong prolate deformation. 
Indeed, HF calculations with the \textsc{ev8} code \cite{bon05} and the SLy4$d$ Skyrme functional \cite{kim97} of the Skyrme EDF \cite{sky56} give a quadrupole deformation parameter $\beta_2\simeq0.4$ \cite{sim04}. 
As mentioned in Sec.~\ref{sec:TDHF}, one advantage of the TDHF theory is that it describes the structure and the reaction mechanisms on the same footing. 
It is then well suited to investigate the role of the orientation of deformed nuclei on reaction mechanisms \cite{sim04,uma06c,sim08,uma08b,gol09,ked10}. 

In the present case, we can then investigate the role of the orientation of the $^{24}$Mg fragments at contact on the outcome of the reaction at $L=36\hb$ and $E_{c.m.}=45.7$~MeV, that is, where the $36^+$ resonance is expected. 
In particular, we can determine  which relative orientations (e.g., tip-tip, tip-side, side-side) lead to the formation of a di-nuclear system. 

\begin{figure}
\sidecaption
\includegraphics[width=7.5cm]{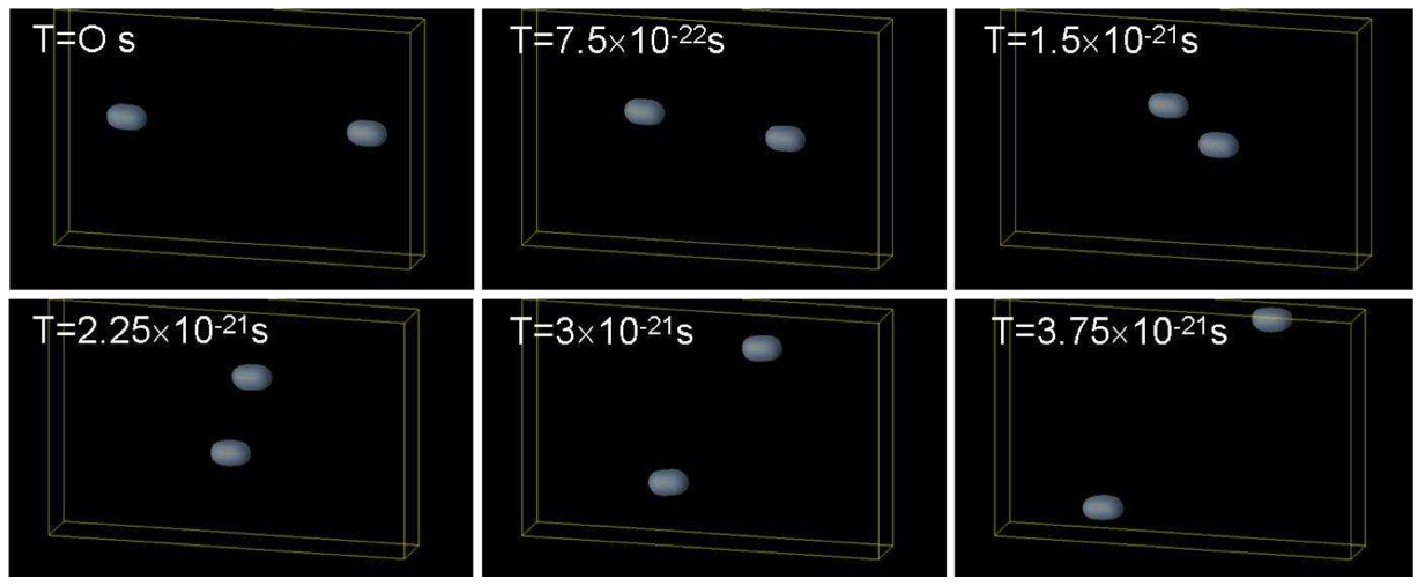}
\caption{TDHF calculation of the reaction $^{24}$Mg+$^{24}$Mg at $L=36\hb$ and $E_{c.m.}=45.7$~MeV. 
The isodensity is plotted at half the saturation density, i.e., $\rho_0/2=0.08$~fm$^{-3}$.
The deformation axes are aligned with the $x$-axis at the initial time of the collision. 
\label{fig:MgMgxx}}
\end{figure}

\begin{figure}
\sidecaption
\includegraphics[width=7.5cm]{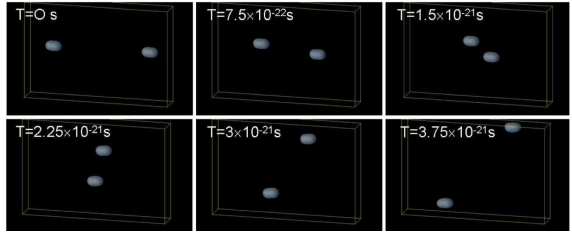}
\caption{Same as Fig.~\ref{fig:MgMgxx} with the deformation axis of the right nucleus aligned with the $y$-axis at initial time. 
\label{fig:MgMgxy}}
\end{figure}

\begin{figure}
\sidecaption
\includegraphics[width=7.5cm]{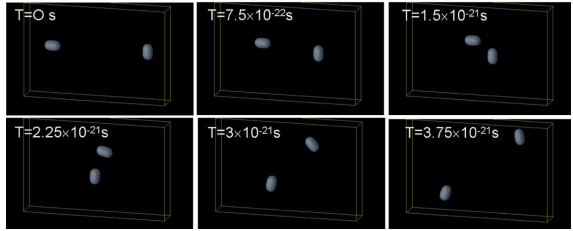}
\caption{Same as Fig.~\ref{fig:MgMgxx} with both deformation axes aligned with the $y$-axis at initial time. 
\label{fig:MgMgyy}}
\end{figure}

Figures~\ref{fig:MgMgxx}, \ref{fig:MgMgxy} and \ref{fig:MgMgyy} present density evolutions of the $^{24}$Mg+$^{24}$Mg system at $L=36\hb$ and $E_{c.m.}=45.7$~MeV for different initial orientations of the nuclei computed with the \textsc{tdhf3d} code using the SLy4$d$ Skyrme functional \cite{kim97}. 
We observe that the contact time between the fragments, if any, is extremely short as compared to the lifetime of the resonance, except when the contact occurs by the tip of the deformed collision partners, as it can be seen in Fig.~\ref{fig:MgMgyy}. 
In the latter case, a di-nuclear system is formed with a lifetime similar or greater than the one of the resonance. 
As a result, it is likely to populate the resonance.
This resonance could then be associated to a molecular state with the two fragments linked by a neck between their tips. 
In particular, it corresponds to a hyper-deformed nucleus as we can see in Fig.~\ref{fig:MgMgyy}. 

To sum up, TDHF calculations provide an insight into the formation of the $36^+$ resonance in $^{24}$Mg+$^{24}$Mg. 
In particular, this resonance may be associated to a highly deformed state of the $^{48}$Cr formed by the two aligned $^{24}$Mg in contact by their tips.

\section{$\alpha$-clustering \label{sec:al-cluster}}

$\alpha$-clusters  play an important role in nuclear structure, both for ground- and excited states, due to the high binding energy of $^4$He \cite{hor10,oer10,gup10,des12,yam12,oer06,fur10,ich11,ebr12}. 
They may also affect reactions. 
For instance, large cross-sections for the $^{7}$Li+$^{208}$Pb$\rightarrow2\al+^{207}$Tl were recently measured \cite{luo11}, whereas this channel has a large negative Q-value. 
$\al$-clustering in the entrance and/or exit channel is also believed to enhance $\al$-transfer in heavy-ion collisions. 
One recent illustration is the formation of $\al$-cluster states in $^{212}$Po with the reaction $^{208}$Pb($^{18}$O,$^{14}$C) \cite{ast10}.
$\al$-clustering is also playing an important role in astrophysical processes such as helium burning in stars \cite{fow86}.

In this section, we present examples of reactions to illustrate the role of $\al$-clusters in nuclear reactions. 
We first investigate the $^{4}$He+$^{8}$Be reaction of astrophysical interests, and, then, we study the survival of an $\al$-cluster in its fusion with a $^{208}$Pb nucleus.

\subsection{Three-$\al$ cluster configurations in $^{4}$He+$^{8}$Be}

\begin{figure}
\sidecaption
\includegraphics[width=7.5cm]{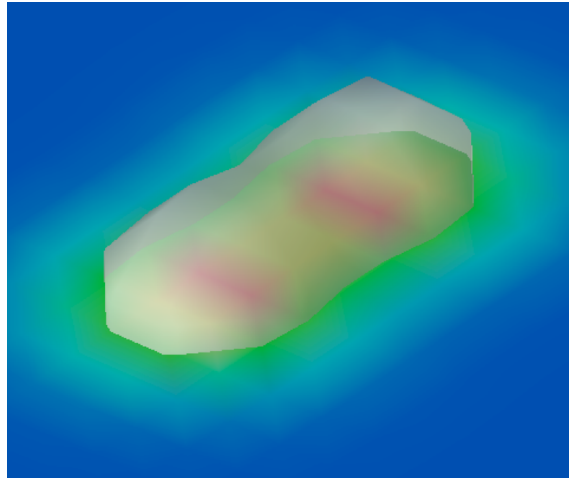}
\caption{HF ground-state density of $^{8}$Be obtained with the SLy4d parametrisation \cite{kim97} of the Skyrme EDF.
\label{fig:dens8Be}}
\end{figure}

The $^{8}$Be ground-state exhibits a 2-$\al$-cluster configuration \cite{hor10,gup10}.
This can be seen in Fig.~\ref{fig:dens8Be} which shows the density profile of $^{8}$Be obtained from a HF calculation. 
In fact, the $^{8}$Be is unbound and decays by the emission of $2\al$ with a lifetime of the order of $\sim10^{-16}$~s. 
This lifetime, however, is long enough to allow the $^{4}$He+$^{8}$Be fusion by radiative capture to occur in stars. 
This reaction is considered as the main source of $^{12}$C production in the universe \cite{fow86}. 

In this reaction, the $^{12}$C is formed in the $0^+_2$ state at $E^*=7.654$~MeV of excitation energy, i.e., just above the 3-$\al$ separation threshold. 
This state, also known as the Hoyle state \cite{hoy54}, is believed to have a strong 3-$\al$ cluster configuration \cite{hor10,des12,yam12}. 
The exact configuration of the $\al$-clusters in the Hoyle state is style under debate.
in particular, it has been suggested that 3-$\al$ linear chains may contribute \cite{mor56,fuj80}. 

\begin{figure}
\sidecaption
\includegraphics[width=7cm]{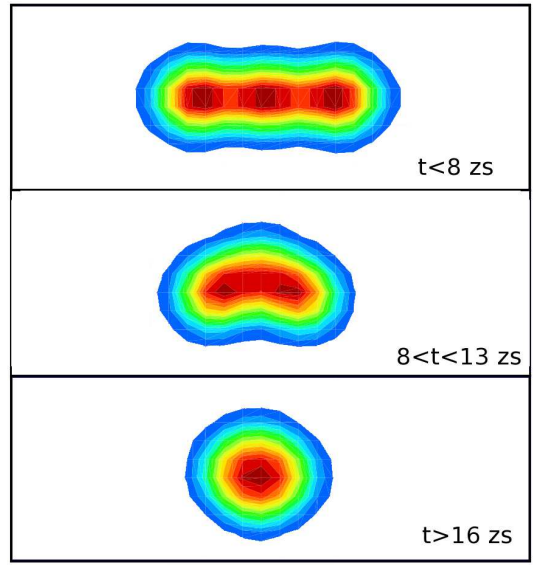}
\caption{Density snapshots of a $^{4}$He+$^{8}$Be reaction at $E_{c.m.}=2$~MeV leading to a 3-$\al$ linear chain (top) and evolving toward a triangular shape (middle) and then toward a more compact shape (bottom). Adapted from Ref.~\cite{uma10a}.
\label{fig:3alpha}}
\end{figure}

Recently, Umar and collaborators have investigated the dynamics of  3-$\al$ linear chains formed in the $^{4}$He+$^{8}$Be reaction using the TDHF formalism \cite{uma10a}.
A 3-$\al$ linear chain is formed by the capture of a $^{4}$He by the tip of the $^{8}$Be in low-energy central collisions. 
Density profiles illustrating different stages of the dynamics of a 3-$\al$ linear chain formed in such a central collision are shown in Fig.~\ref{fig:3alpha}. 
The linear chain (top of Fig.~\ref{fig:3alpha}) is present during a long time (up to 8~zs). 
During this time, the $\al$-clusters present a longitudinal vibrational mode \cite{uma10a}. 
It is interesting to note that the clusters remain for such a long time, while the underlying formalism is a mean-field model of independent particles, i.e., without imposing the presence of such clusters in the wave-function. 

For longer times, the chain becomes unstable due to the appearance of a bending motion favouring the formation of triangular shapes (middle of Fig.~\ref{fig:3alpha}). 
The clusters still encounter some vibrations in this mode, with the center cluster oscillating perpendicular to the left and right clusters. 
This vibration mode last for another $\sim4$~zs with almost no damping before a more contact shape is formed (bottom of Fig.~\ref{fig:3alpha}).

\begin{figure}
\sidecaption
\includegraphics[width=7.5cm]{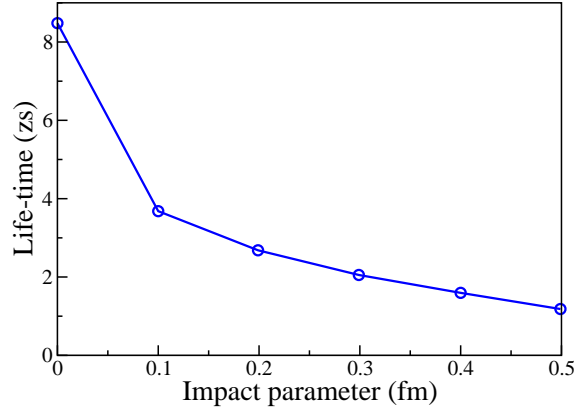}
\caption{Lifetime of the 3-$\al$ chain as a function of the impact parameter for the $^{4}$He+$^{8}$Be reaction at $E_{c.m.}=2$~MeV.  Adapted from Ref.~\cite{uma10a}.
\label{fig:3alpha_t}}
\end{figure}

The role of impact parameter has also been studied by Umar {\it et al.} in Ref.~\cite{uma10a}. 
A decrease of the lifetime of the linear chain with the impact parameter was observed, as reported in Fig.~\ref{fig:3alpha_t}. 
We note that, although non-central collisions do not favour long lifetime of linear chains, they are stable enough to survive more than 1~zs as long as the collision occurs with the tip of $^8$Be and for impact parameters not exceeding 0.5~fm. 
These times are of the same order of magnitude, if not larger, than typical lifetimes of di-nuclear systems formed in near-barrier heavy-ion collisions.

To conclude, relatively long lifetimes of few zs are observed for linear chains of 3-$\al$ clusters formed in $^{4}$He+$^{8}$Be within the TDHF approach which does not assume {\it a priori} cluster components in the wave-function. 
The dynamics of these structures exhibit complex vibrational modes based on oscillations of the $\al$-cores. 
In fact, similar vibrational modes have  been found with the fermionic molecular dynamics approach by Furuta and collaborators \cite{fur10}. 
In particular, possible strong effects on the vibrational response functions of light nuclei have been noticed.

\vspace{1cm}

\subsection{Survival of $\al$-clusters in $^{4}$He+$^{208}$Pb near-barrier fusion}

The previous section emphasises the survival of $\al$-clustering after a capture process in a light system.
$\al$-clustering is not limited, however, to light nuclei.
Indeed, the well-known $\al$-radioactivity, which occurs essentially in heavy nuclei, is another form of $\al$-clustering. 
In addition, the reverse process of capture of an $\al$ by a heavy nucleus in a transfer or a fusion reaction may form excited states interpreted as nucleus+$\al$ molecules. 
For instance, new excited states have been recently observed in $^{212}$Po that are interpreted as $\al+^{208}$Pb configurations \cite{ast10}. 

The dynamics of the $\al+^{208}$Pb system after capture of the $\al$ by the heavy partner has been investigated with an early TDHF code in Ref.~\cite{san83}. 
However, these calculations were performed with a simplified Skyrme functional. In particular, they did not include the spin-orbit interaction which is known to be crucial for a proper description of nuclear reactions \cite{uma86,mar06,uma06a}. 
In fact, the spin-orbit interaction and the difference between proton and neutron mean-fields may induce a ''dissolution'' of an $\al$-particle entering the mean-field of a collision partner \cite{iwa08}. 

\begin{figure}
\begin{center}
\includegraphics[width=4cm]{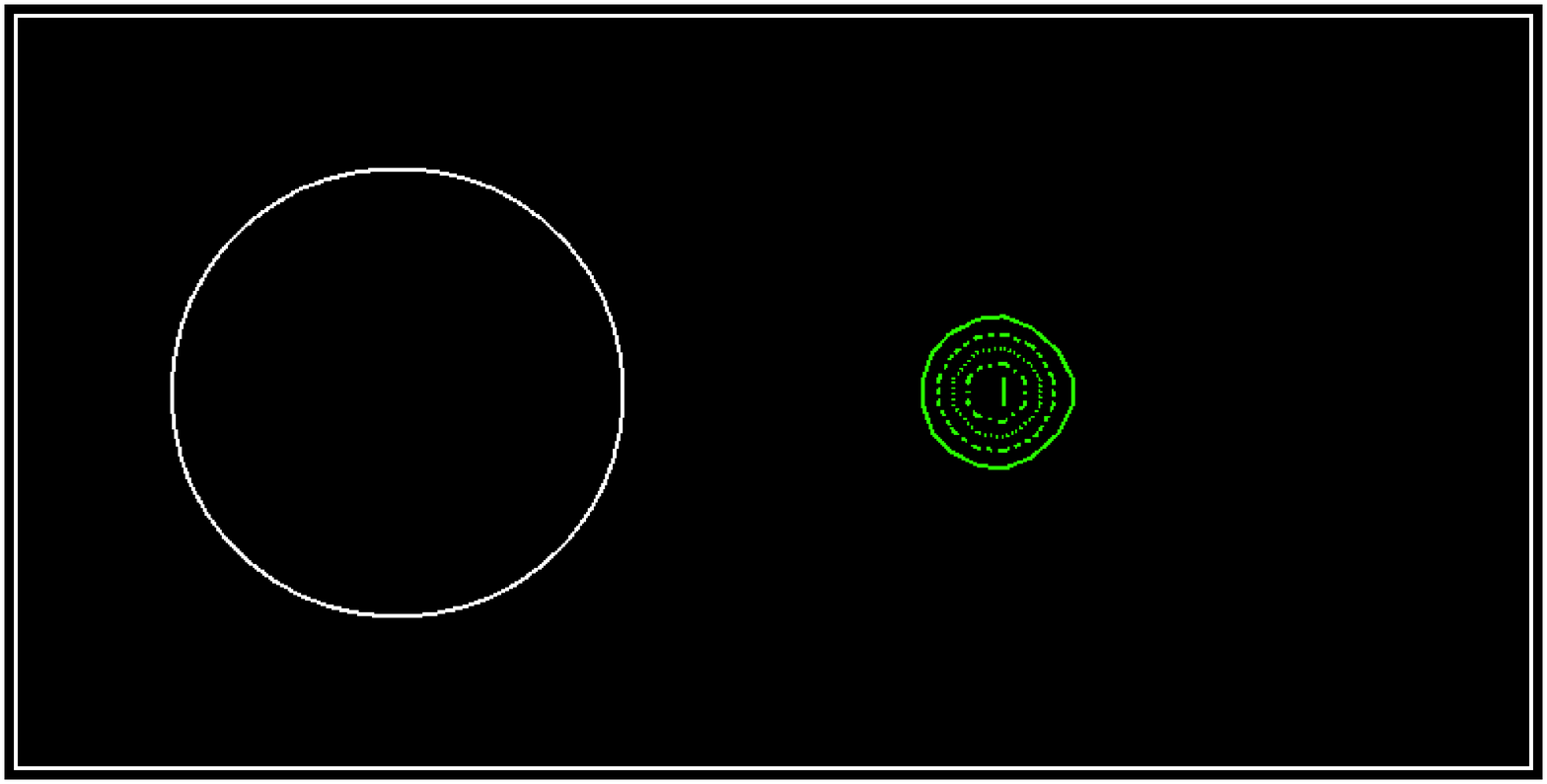}\\
\includegraphics[width=4cm]{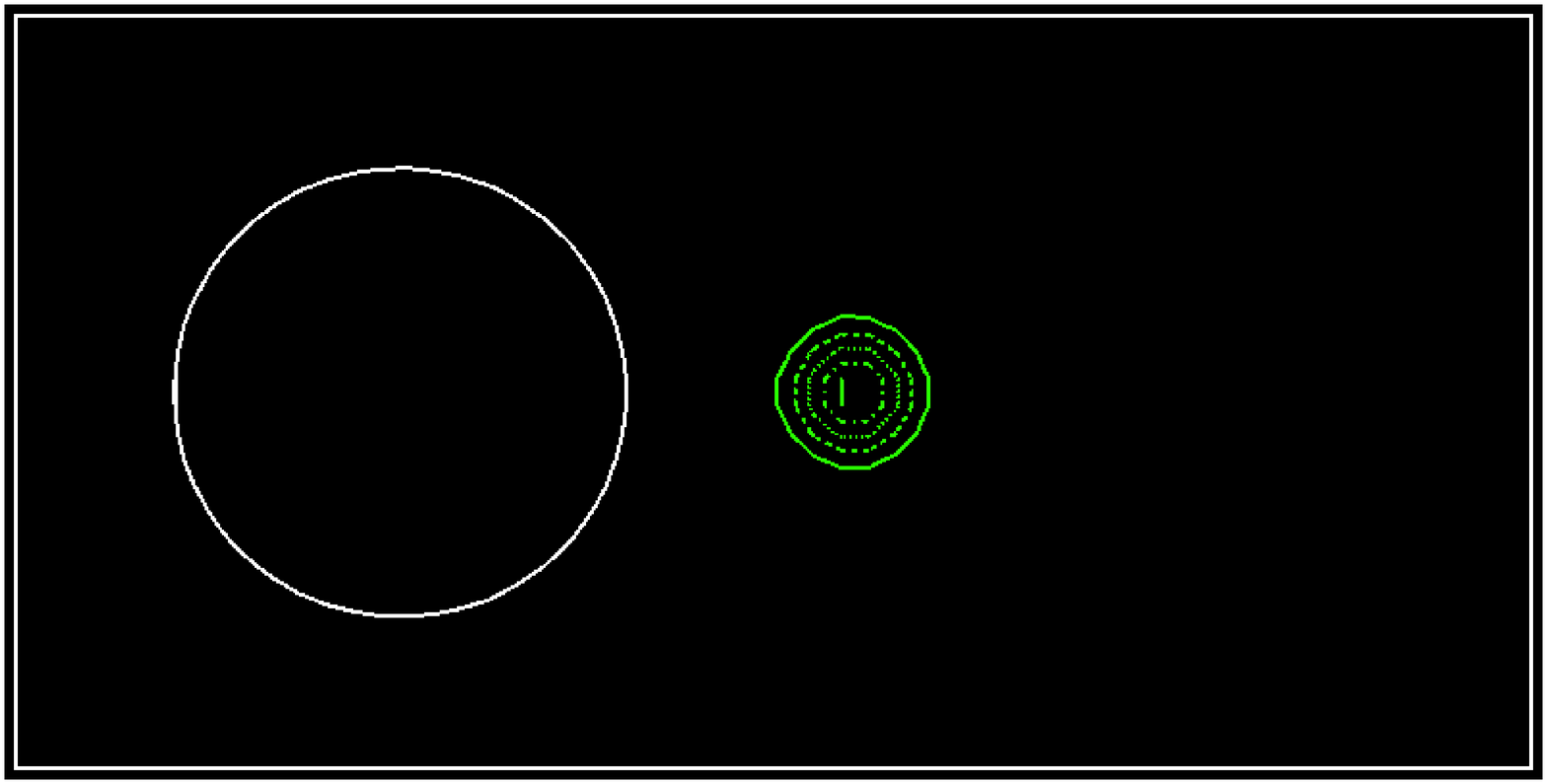}\\
\includegraphics[width=4cm]{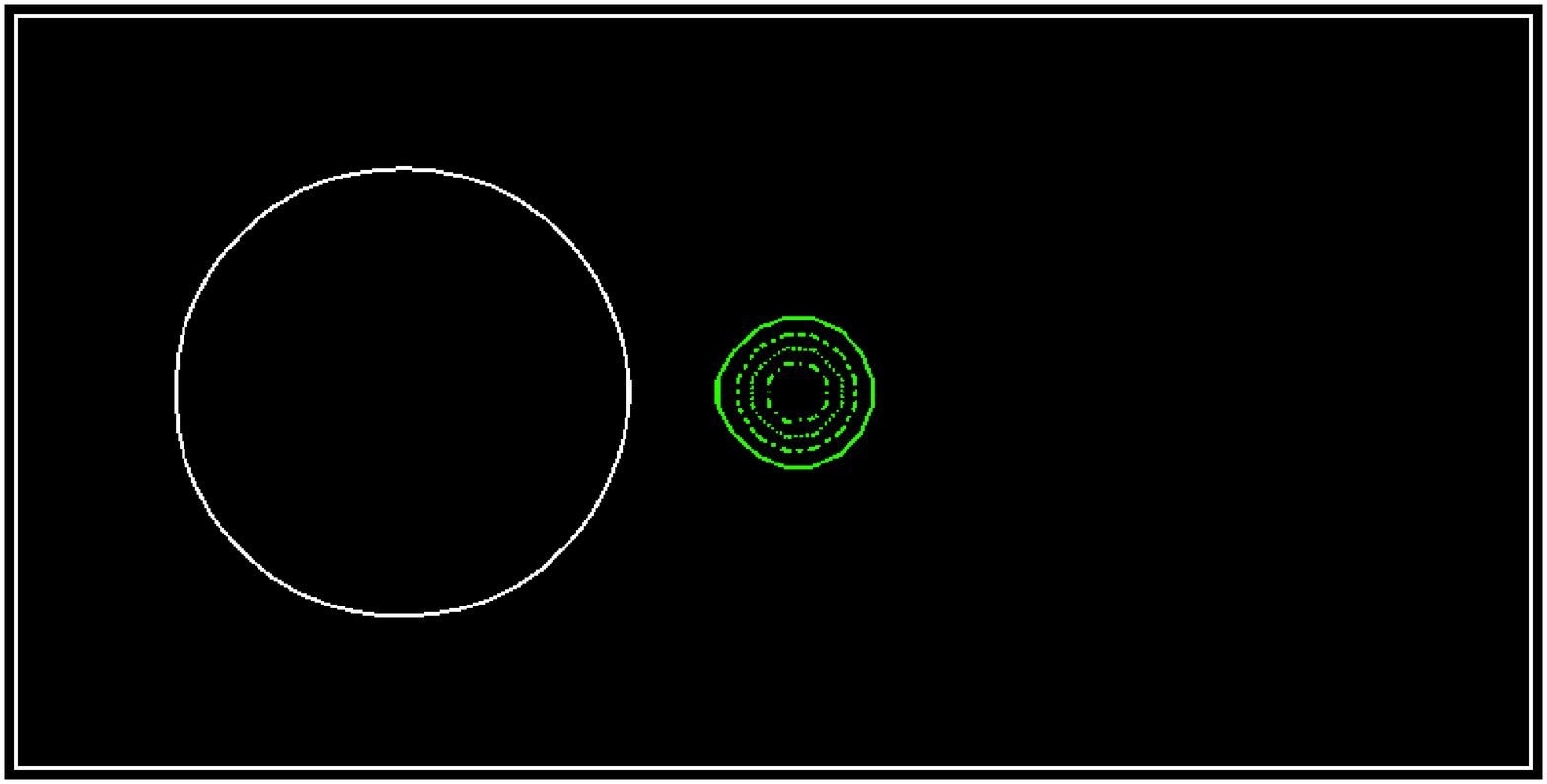}\\
\includegraphics[width=4cm]{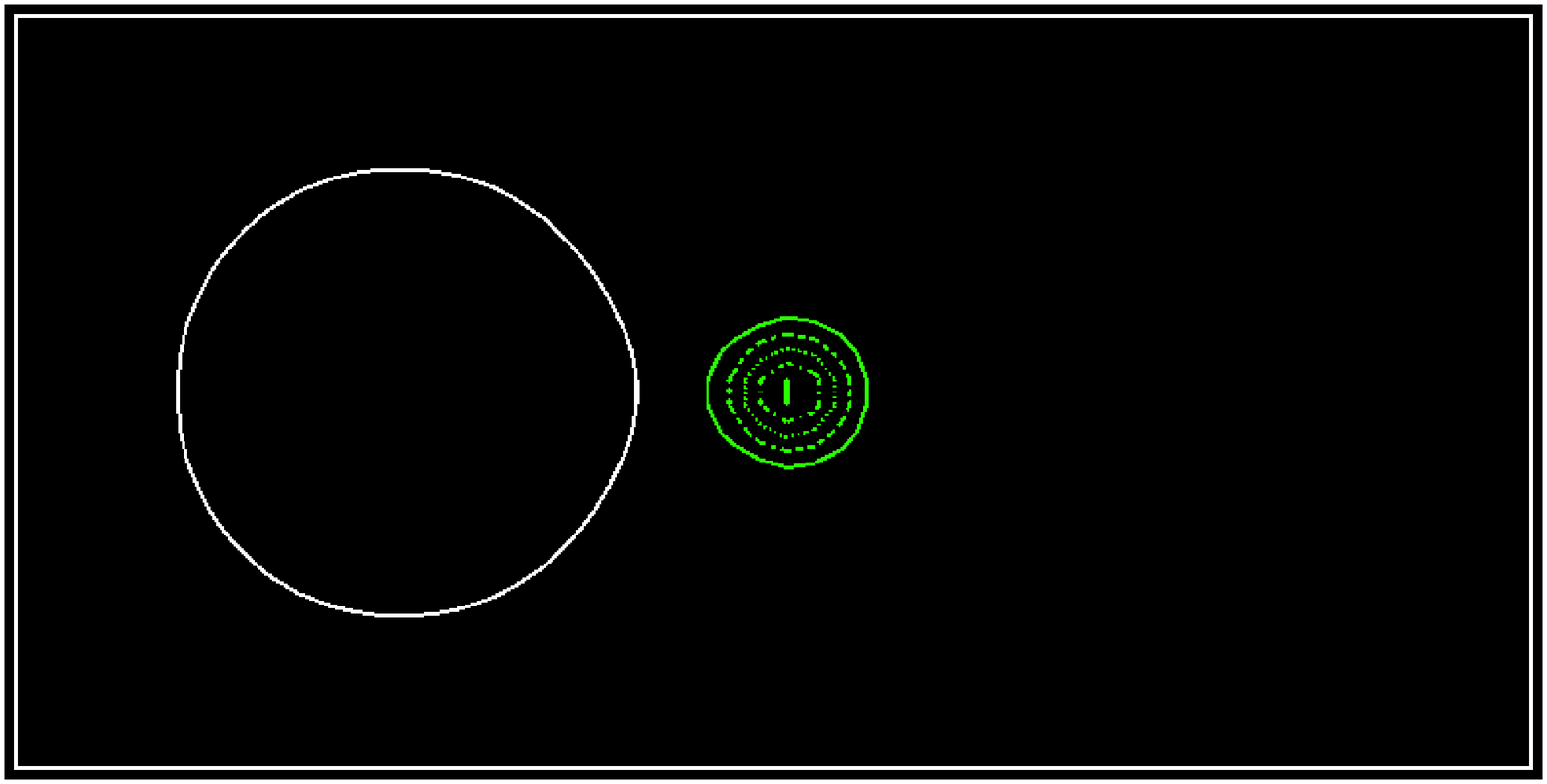}\\
\includegraphics[width=4cm]{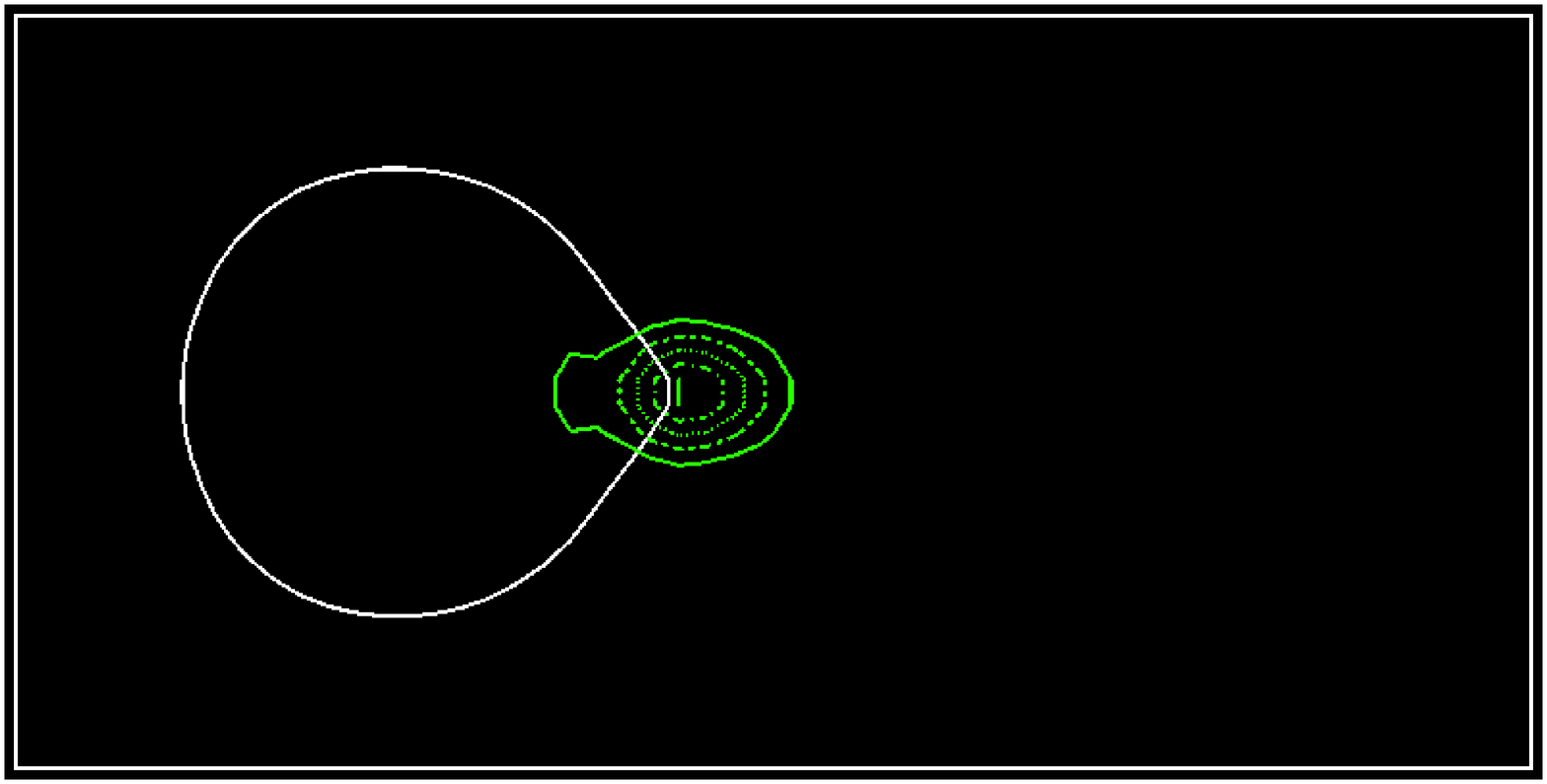}\\
\includegraphics[width=4cm]{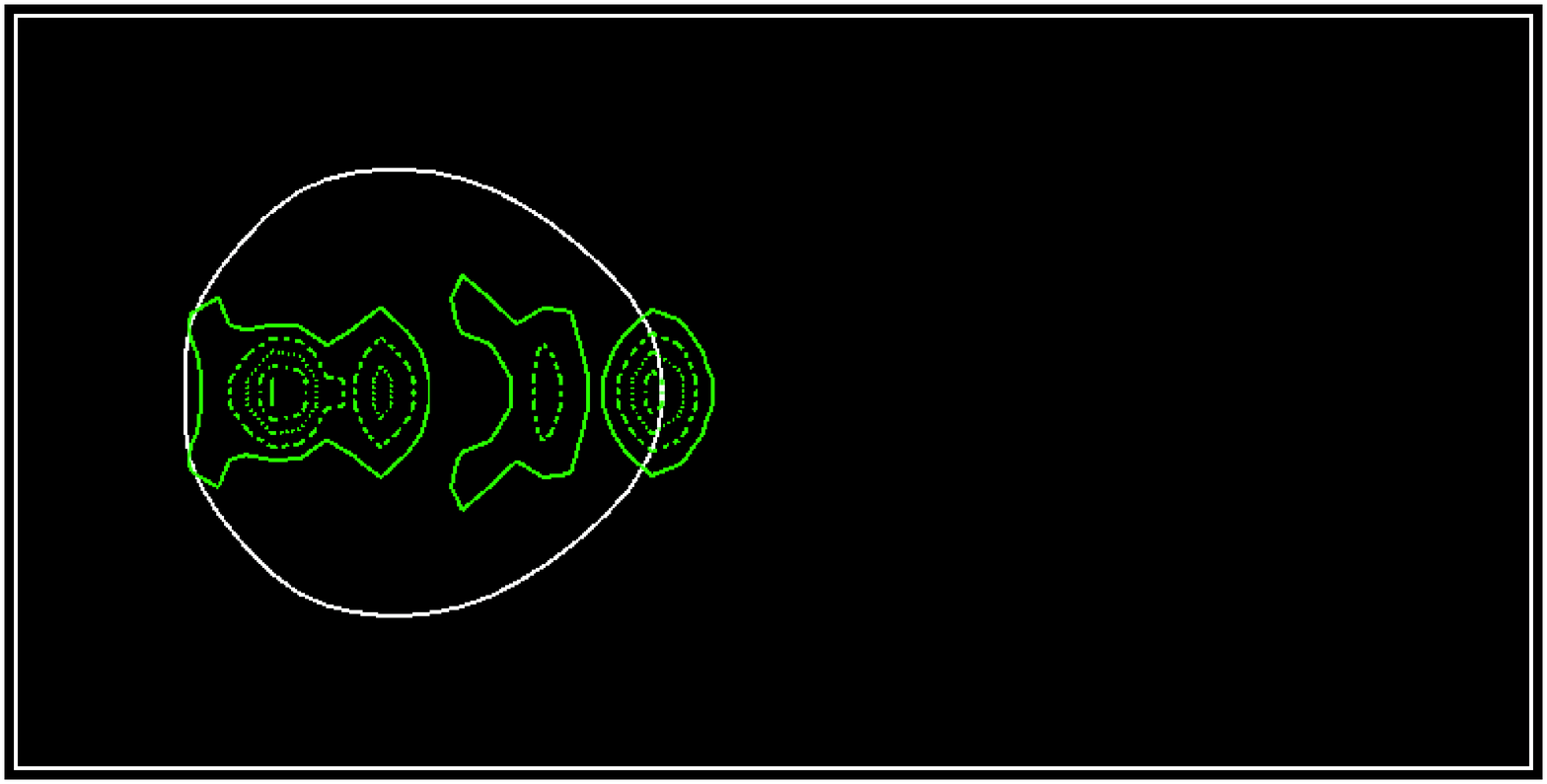}\\
\includegraphics[width=4cm]{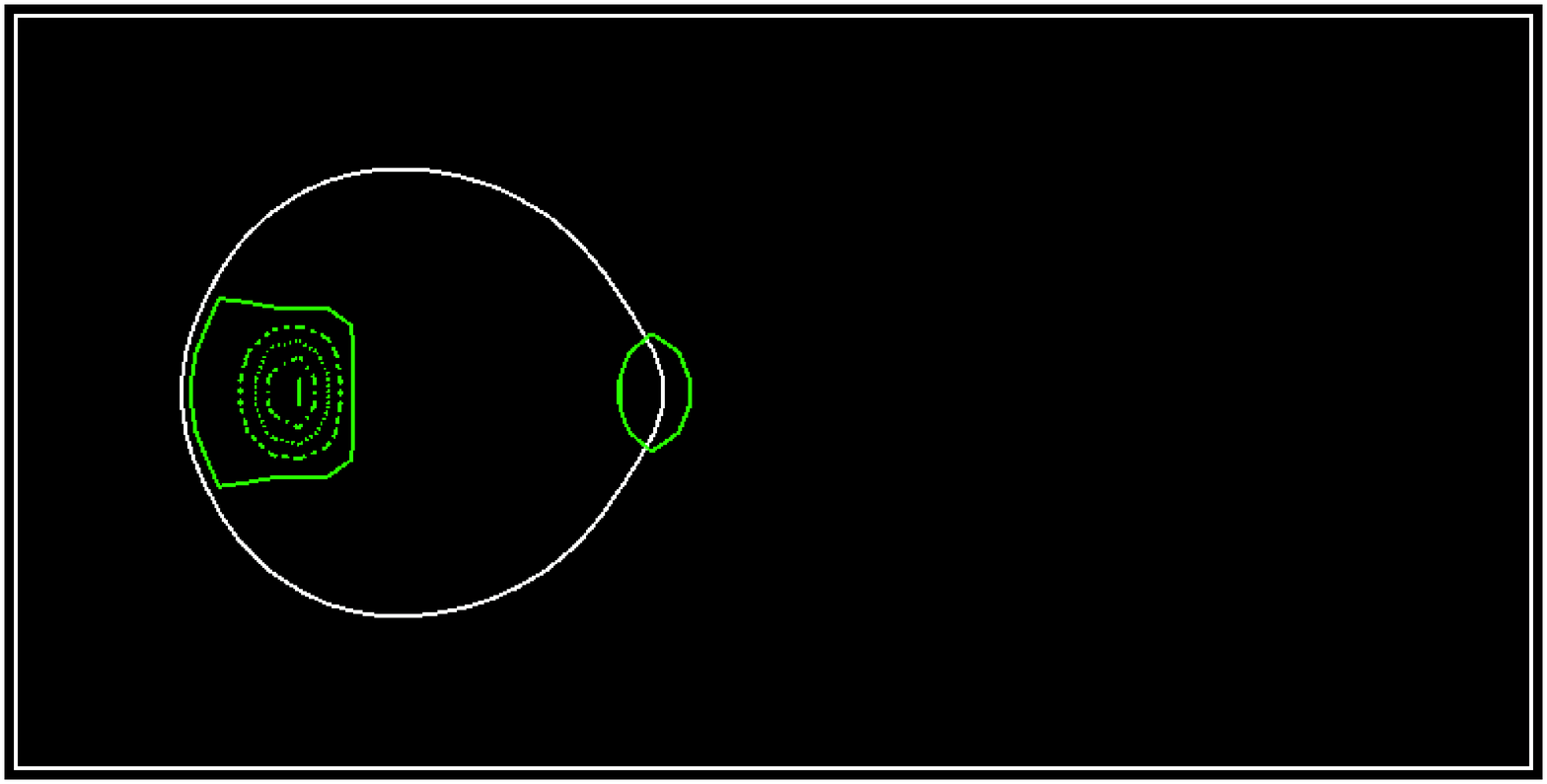}\\
\includegraphics[width=4cm]{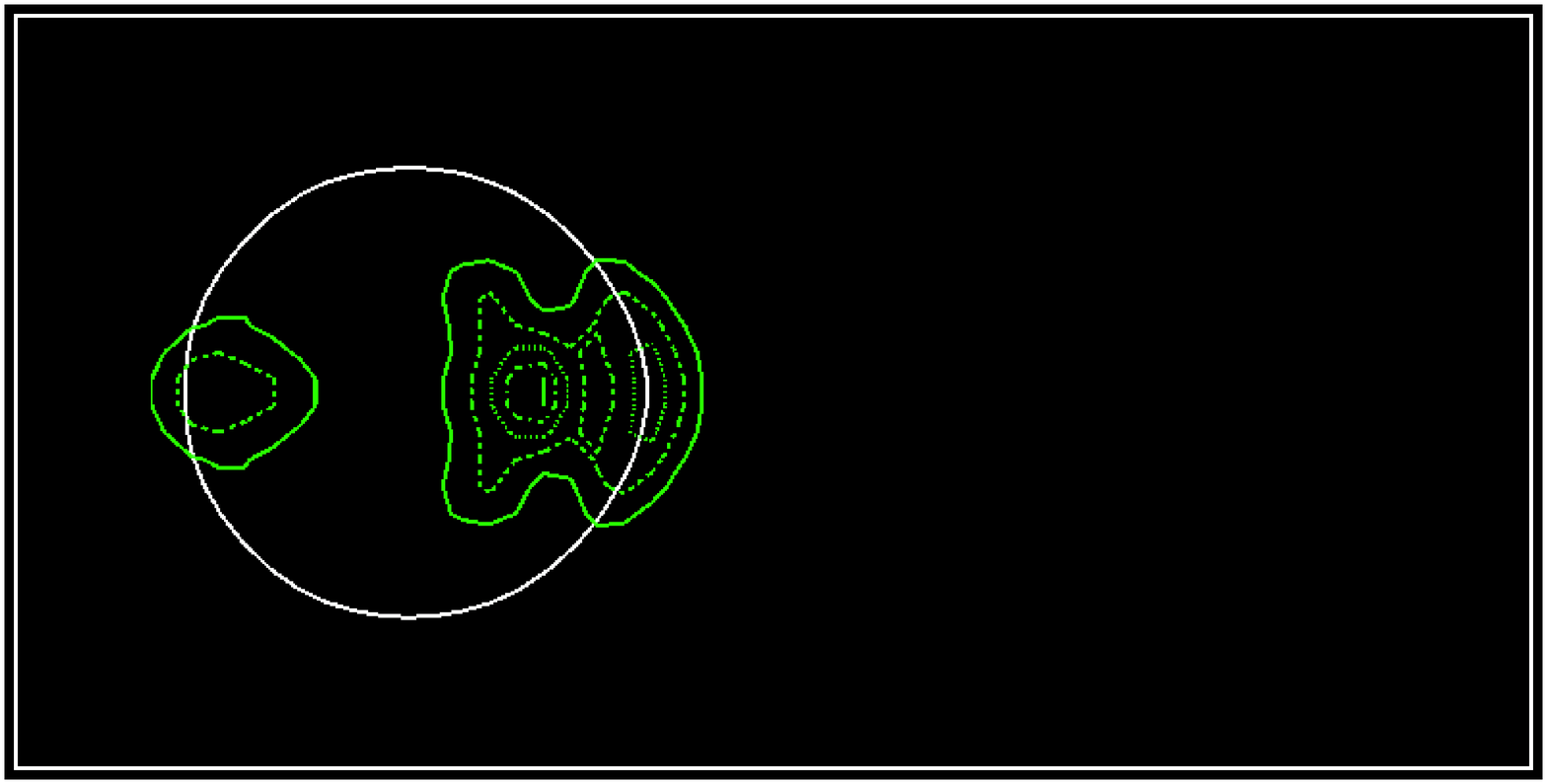}\\
\includegraphics[width=4cm]{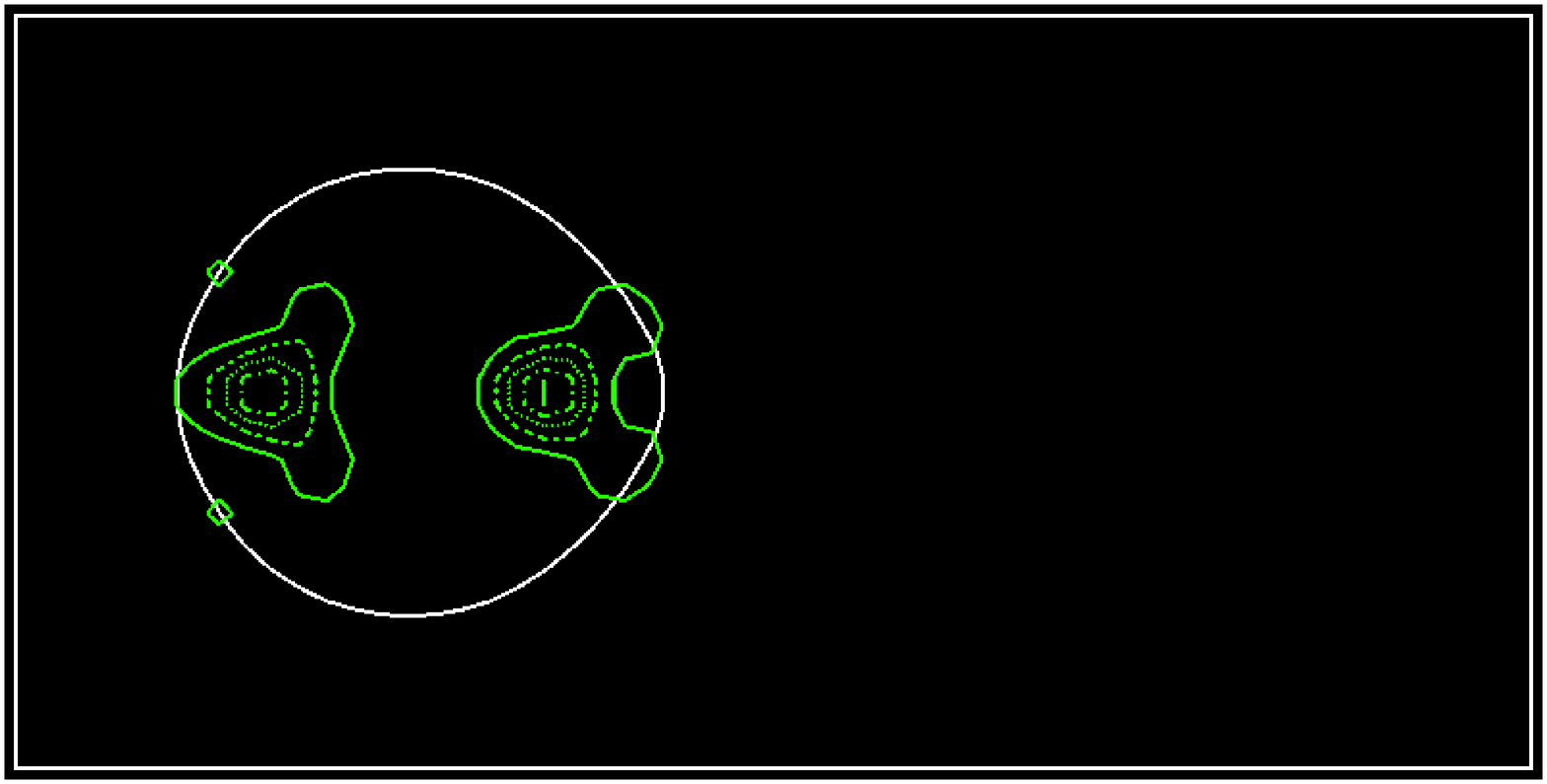}\\
\includegraphics[width=4cm]{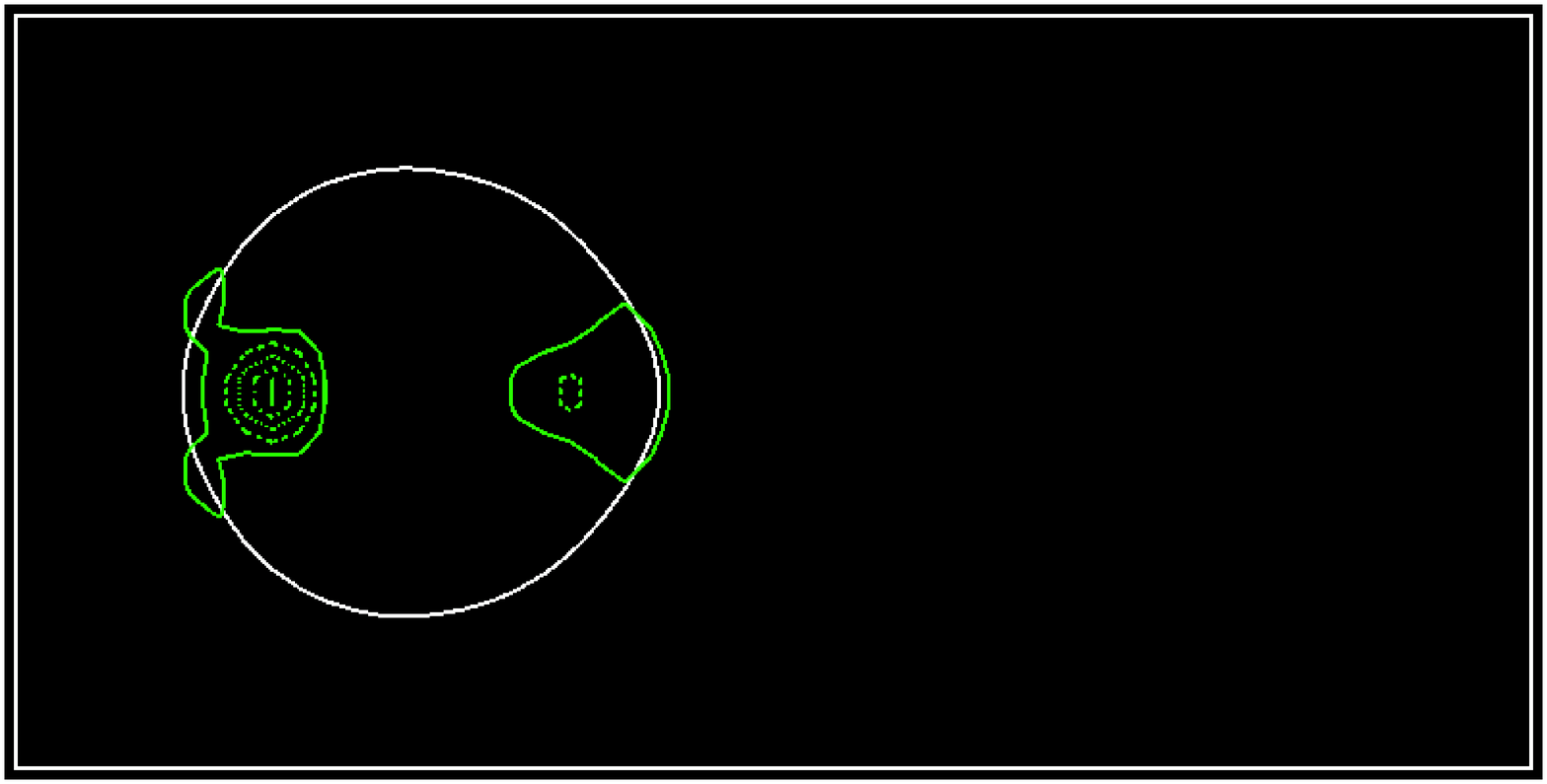}
\end{center}
\caption{Snapshots of the density profile in the $^4$He+$^{208}$Pb central collision at $E_{c.m.}=19.8$~MeV.
Wave-functions initially belonging to the $^4$He are shown with green contours. A single isodensity at half the saturation density (white solid line) is shown for the wave-functions initially belonging to the $^{208}$Pb fragment. 
\label{fig:HePb}}
\end{figure}

To test this idea, three-dimensional TDHF calculations with the \textsc{tdhf3d} code and the SLy4$d$ Skyrme functional \cite{kim97} including spin-orbit terms have been performed. 
The $^4$He+$^{208}$Pb central collision at $E_{c.m.}=19.8$~MeV, i.e., slightly above the Coulomb barrier, is considered. 
The evolution of the density associated to the single-particle wave-functions initially forming the $^4$He projectile is shown with green contours in Fig.~\ref{fig:HePb}. 
A clear dissolution of the $\al$-particle in at least two components is observed as soon as it enters the mean-field of the heavy partner, represented by a single isodensity at half the saturation density (white solid line in Fig.~\ref{fig:HePb}). 

To conclude, the correlations responsible for the survival of $\al$-clustering in the previous study of the  $^4$He+$^8$Be system are not strong enough, at the mean-field level, to enable similar effects in a $^4$He+$^{208}$Pb central collision at the barrier. 
Indeed, the $2p$-$2n$ spatial correlations in the $^{4}$He fragment are lost in the mean-field of the heavy partner.
The experimentally observed $\al$-cluster states in the $^{212}$Po nucleus \cite{ast10} may either be due to beyond-mean-field correlations, or to an $\al$-cluster located at the surface and orbiting around the heavy core. 
To test the latter idea, the present calculations should be repeated above the barrier and around the grazing angle to investigate possible long-lived $\al$-cluster configurations. 

\section{Transfer in heavy-ion collisions \label{sec:transfer}}

Transfer reactions are highly sensitive to cluster effects. 
This is illustrated in the $\al$-transfer experiment of Ref.~\cite{ast10} discussed above.
Other types of clusters may also be transferred in heavy-ion collisions, such as pairs of protons and neutrons \cite{oer01}.
These types of multi-nucleon transfer reactions are in competition with the sequential transfer, i.e., a transfer of independent nucleons. 

Theoretically, transfer probabilities in microscopic approaches are obtained from a particle number projection technique. 
First, this technique is described.
It is then applied at the TDHF level  to estimate sequential transfer probabilities in the $^{16}$O+$^{208}$Pb system. 
The latter are used as a reference to interpret experimental data in terms of cluster transfer. 
Finally, pairing vibrations, which may be excited in pair transfer reactions, are studied with the TDHFB approach. 

\subsection{Particle number projection technique}

Transfer probabilities have been extracted at the TDHF level \cite{koo77,sim10b} thanks to a projection onto a good particle number technique. 
This technique is standard in beyond-mean-field models for nuclear structure subject to pairing correlations~\cite{rin80,bender03}.
Here, it is applied on the outgoing fragments to determine their proton and neutron number probabilities. 

It is possible to extract the component of the wave function associated to a specific transfer channel using a particle number projector onto $N$ protons or neutrons in the $x>0$ region where one fragment is located at the final time, the other one being in the $x<0$ region.
Such a projector is written~\cite{sim10b}
\oeq
\oP_R(N)=\frac{1}{2\pi}\int_0^{2\pi} \stb \d \tet \stf e^{i\tet(\oN_R-N)},
\label{eq:projector}
\ceq
where
\oeq
\oN_R = \sum_{s} \sdf \int \stb \d \vr \stf \oad(\vr s) \sdf \oa(\vr s) 
\sdf \Theta(x)
\label{eq:NG}
\ceq
counts the number of particles in the $x>0$ region ($\Theta(x)=1$ if $x>0$ and $\Theta(x)=0$ elsewhere).
Isospin is omitted to simplify the notation. 

The projector defined in Eq.~(\ref{eq:projector}) can be used to compute the probability to find $N$ nucleons in $x>0$ in the final state $\kfi$,
\oeq
\left|\oP_R(N)\kfi\right|^2=\frac{1}{2\pi}\int_0^{2\pi} \stb \d \tet \stf e^{-i\tet{N}}\bfi\phi_R(\tet)\>,
\label{eq:proba}
\ceq
where $|\phi_R(\tet)\>=e^{{i\tet\oN_R}}\kfi$ 
 represents a rotation of $\kfi$ by a gauge angle $\tet$ 
in the gauge space associated to the particle number degree of freedom in $x>0$.
Note that $|\phi_R(\tet)\>$ is an independent particle state. 
The last term in Eq.~(\ref{eq:proba}) is then the determinant of the matrix of the occupied single particle state overlaps \cite{sim10b}:
\oeq
\bfi\phi_R(\tet)\>=\det (F)
\ceq
with
\oeq
F_{ij}= \sum_{s} \int \stb\d \vr \sdf{\az_i^s}^*(\vr) {\az_j^{s}}(\vr) e^{i\tet\Theta(x)}.
\ceq
The integral in Eq.~(\ref{eq:proba}) is discretised using $\tet_n=2\pi{n}/M$ with the integer $n=1\cdots{M}$.
Choosing $M=300$ ensures numerical convergence for the $^{16}$O+$^{208}$Pb system. 

\subsection{Sub-barrier transfer in $^{16}$O+$^{208}$Pb\label{sec:transOPb}}

We mentioned in Sec.~\ref{sec:al-cluster} the $\al$-transfer reaction $^{208}$Pb($^{18}$O,$^{14}$C) populating $\alpha$-cluster states in $^{212}$Po \cite{ast10}.
One could expect a similar $\al$-transfer mechanism to dominate the $^{208}$Pb($^{16}$O,$^A$C) reaction channels below and around the barrier \cite{vri75,has79,tho89}. 
Indeed, $^{16}$O is a good candidate for $\al$-clustering, and $\al$-condensates \cite{fun08} as well as linear $\al$-chains \cite{ich11} have been predicted.
Experimental indications of $\al$-clustering in $^{16}$O have also been reported \cite{whe11,abd03}.
Recent experimental data \cite{eve11} showed, however, that the most probable carbon isotope formed in $^{16}$O+$^{208}$Pb is $^{14}$C, indicating a dominance of two-proton transfer against $\al$-transfer. 

\begin{figure}
\sidecaption
\includegraphics[width=7.5cm]{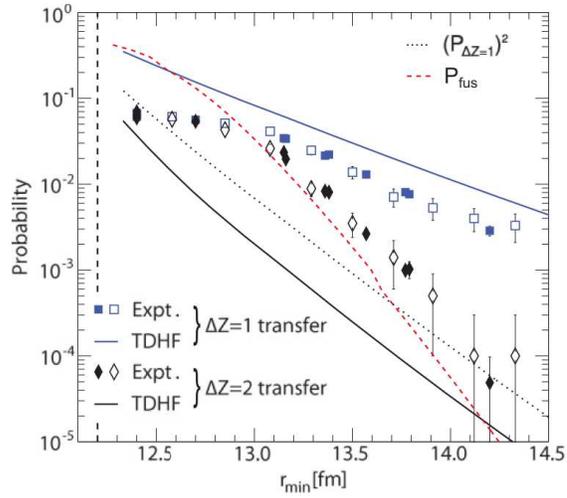}
\caption{Charge transfer probability in $^{16}$O+$^{208}$Pb. TDHF calculations of the $\Delta{Z}=1$ (1p stripping) and $\Delta{Z}=2$ (2p stripping) transfer probabilities as a function of the distance of closest approach $r_{min}$ are shown by the full curves. Experimental data are shown for $\Delta{Z}=1$ (squares) and $\Delta{Z}=2$ (diamonds). The vertical dashed line indicates the average barrier radius. The dotted curve shows the square of the TDHF $\Delta{Z}=1$ transfer probabilities. The red dashed curve shows the sub-barrier fusion probability as calculated with the coupled-channels code CCFULL. From Ref.~\cite{eve11}.
\label{fig:transfer}}
\end{figure}

TDHF calculations \cite{sim10b} have been performed with the \textsc{tdhf3d} code and the SLy4$d$ Skyrme functional \cite{kim97} to estimate the probability for production of nitrogen ($\Delta{Z}=1$) and carbon ($\Delta{Z}=2$) in $^{16}$O+$^{208}$Pb. The results are shown in Fig.~\ref{fig:transfer} (solid lines) as a function of the distance of closest approach for Coulomb trajectories
\oeq
r_{min}=\frac{Z_1Z_2e^2}{2E_{c.m.}}\(1+\mbox{cosec}\frac{\theta_{c.m.}}{2}\).
\ceq
We observe that these probabilities decrease exponentially with increasing $r_{min}$, as expected for quasi-elastic transfer \cite{cor09}.

Due to the independent particle approximation, the TDHF approach is only able to describe sequential multi-nucleon transfer, i.e., neglecting cluster correlations. 
The probability for sequential transfer of two nucleons is sometimes approximated by the square of the one-nucleon transfer probability. 
This leads, however, to an overestimation of the two-proton sequential transfer in $^{16}$O+$^{208}$Pb, as can be seen in Fig.~\ref{fig:transfer} (compare dotted and solid black lines). 
In fact, the above approximation is valid when a large number of particles are available for transfer toward states with large degeneracies. 
In general, this criterion  is not fulfilled due essentially to the relatively small number of single-particle states around the Fermi level. 
Indeed,  the latter have the smallest binding energies, and, then, the largest transfer probabilities.
In the $^{16}$O case, for instance, transfer is dominated by the $1p_{1/2}$ single particle states. 
Quantum microscopic approaches such as the TDHF theory are then needed to estimate correct sequential multi-nucleon transfer probabilities. 

Comparison with experimental data in Fig.~\ref{fig:transfer} indicates that to assume pure sequential transfer, as in TDHF calculations, leads to an overestimation of one-proton transfer probabilities by a factor $\sim2$, and to an underestimation of two-proton transfer probabilities by approximatively one-order of magnitude. 
This is interpreted as a strong cluster effect in $^{16}$O+$^{208}$Pb charge-transfer \cite{sim10b,eve11}. 
Indeed,  correlations such as proton pairing or $\al$-clustering favour the transfer of two protons, while they reduce the probability for transferring only one proton. 
In fact, it is shown in Ref.~\cite{eve11} that pairing correlations dominate the $\Delta{Z}=2$ channel over $\al$-clustering. 

It is interesting to note that at large distances of closest approach, i.e., $r_{min}>13$~fm, the sum of the $\Delta{Z}=1$ and 2 channels is rather well reproduced by the TDHF calculations \cite{sim10b}. 
Closer to the barrier, however, sub-barrier fusion dominates, as shown by the coupled-channels calculations with the \textsc{ccfull} code \cite{hag99}. 
As a result, the experimental transfer probabilities are reduced at the barrier, inducing a deviation from the exponential dependence observed at larger distances. 
One drawback of the TDHF approach is that it does not enable tunnelling of the many-body wave-function.
Thus, sub-barrier fusion is not included in TDHF calculations, inducing an overestimation of the total transfer probabilities close to (but below) the barrier. 

To sum up, the TDHF approach provides a good estimate of sequential transfer probabilities well below the barrier. 
These sequential transfer probabilities can be used as a benchmark to compare with experimental data in order to emphasise the role of cluster correlations on the transfer mechanism. 

\subsection{Pairing vibrations\label{sec:pairvib}}

It is possible to include pairing correlations in the mean-field dynamics by considering quasiparticle vacua instead of independent particle states. 
This leads to the TD-BCS model when only pairs between time-reversed single-particle states are considered, or, more generally, to the time-dependent Hartree-Fock-Bogoliubov (TDHFB) formalism \cite{bla86}. 
Numerical applications are now possible thanks to the recent development of TD-BCS \cite{eba10,sca12} and TDHFB \cite{ave08,ste11,has12} codes.

A proper description of pairing dynamics is crucial to investigate the evolution of nuclei produced by a pair transfer mechanism \cite{rip69,oer01,kha04,pll11,shi11,gra12}. 
Indeed, pair transfer reactions are a good tool to excite the so-called ''pairing vibrations'' \cite{boh75,rin80,bes66,rip69,oer01}. 
Pairing correlations are then expected to induce a collectivity which manifests itself as an increase of transition amplitude toward these states. 

Recent studies of pairing vibrations have been performed at the QRPA level \cite{kha04,pll11}, that is, in the linearised version of TDHFB \cite{bla86,rin80}. 
Here, we discuss a similar study with a real-time description of pairing vibrations excited in two-neutron transfer reactions in $^{44}$Ca with a fully self-consistent TDHFB code \cite{ave08,ave09b,ave10}. 
Applications to other systems including oxygen, calcium, and tin isotopes can be found in Refs.~\cite{ave08,ave10,ave09}.

Starting with an even-even nucleus ground-state with $A$ nucleons and spin-parity $0^+$, and assuming a $\Delta L=0$ direct pair transfer reaction, pair vibration states with $J^\pi=0^+$ are populated in the $A+2$ (pair addition) and/or $A-2$ (pair removal) nuclei. 
The transfer process is simulated by an initial excitation generated by a boost with a Hermitean pair-transfer operator~\cite{bes66}
\begin{eqnarray}
\oF = 
\int d\mathbf{r} \, f(r) \left( \oad_{\mathbf{r},\downarrow} 
\oad_{\mathbf{r},\uparrow} 
+ \oa_{\mathbf{r},\uparrow} \oa_{\mathbf{r},\downarrow} \right),
\label{eq:pairtrans}
\end{eqnarray}
where the arrows label the spin of the single-particles (we omit the isospin to simplify the notation). 
In the present application, $f(r)$ is a Fermi-Dirac spatial distribution containing the nucleus and cutting at 4~fm outside the nucleus. 
Its role is to remove unphysical high energy modes associated to pair creation outside the nucleus.

\begin{figure}
\sidecaption
\includegraphics[width=7.5cm]{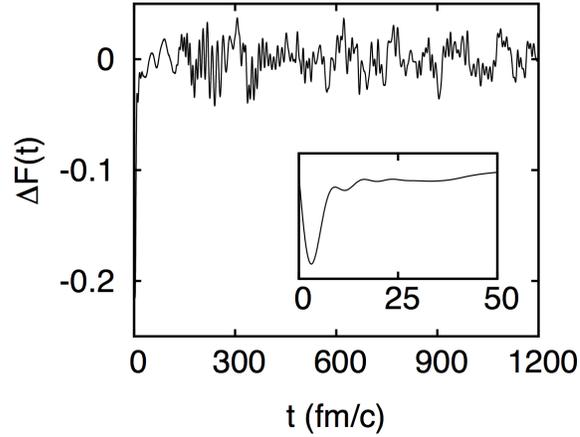}
\caption{Evolution of $\<F\>(t)$ after a pair transfer type excitation on $^{44}$Ca. The inset shows the same quantity at early times. Adapted from Ref.~\cite{ave09}. \label{fig:pairvib}}
\end{figure}

The evolution of the system after the boost is obtained with the \textsc{tdhfbrad} code \cite{ave08}.
This code solves the TDHFB equation in spherical symmetry with a full Skyrme EDF and density-dependent pairing effective interaction. 
The linear response of $\<\oF\>(t)$ after a boost excitation is shown in Fig.~\ref{fig:pairvib} for a $^{44}$Ca initial ground-state.
The SLy4 parametrisation of the Skyrme EDF is used \cite{cha98}, together with a ''volume'' pairing effective interaction of the form $\tilde{t}_0\delta(\ovr_1-\ovr_2)$ with $\tilde{t}_0=-187$~MeV.fm$^{-3}$ and a cut-off energy of 80~MeV in the quasi-particle spectrum to avoid ultra-violet divergence.  
See Refs.~\cite{ave09,kha09,ave10} for a discussion on the role of the form of the pairing functional on the excitation of pairing vibrations. 

\begin{figure}
\sidecaption
\includegraphics[width=7.5cm]{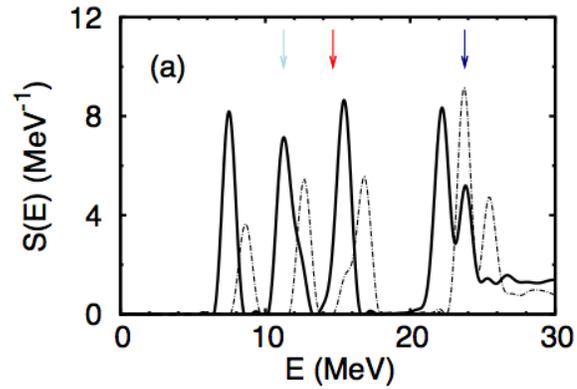}
\caption{TDHFB strength function (solid line) associated to the evolution of $\<F\>(t)$ in Fig.~\ref{fig:pairvib}. The unperturbed spectrum (dotted line) is also shown. 
The arrows indicate pair removal transitions from deep hole states. 
Adapted from Ref.~\cite{ave09}.  
\label{fig:TFpairvib}}
\end{figure}

The apparent chaotic behaviour of $\<\oF\>(t)$ in Fig.~\ref{fig:pairvib} is due to the simultaneous excitation of several states. 
A simple Fourier analysis can be used to extract the energy and relative contributions of these states to the time evolution of $\<\oF\>(t)$. 
The resulting strength function is plotted in Fig.~\ref{fig:TFpairvib} with a solid line. 
Both pair additional and pair removal (indicated by the arrows) modes are present. 
The unperturbed strength function (dashed line) obtained by removing the self-consistency of the generalised mean-field is also shown. 
Overall, an increase of the strength is observed due to the dynamical pairing residual interaction present in TDHFB, but neglected in the unperturbed response. 
This increase of the strength is a signature for collective motion, indicating that several quasi-particles participate to the vibrational modes. 
In addition, this residual interaction lowers the transition energies due to its attractive nature.
These characteristics of pairing vibrations are in agreement with previous observations with the QRPA model \cite{kha04}.

To conclude, collective pairing vibrations may be excited in pair-transfer reactions. 
These reactions are then a unique probe to investigate the dynamics of pairs of nucleons in nuclei, in particular their vibrational modes. 
Pairing vibrations have been studied at the TDHFB level with spherical symmetry. 
The recent development of a three-dimensional TDHFB code \cite{ste11} might enable similar studies in deformed nuclei and for $L\ne0$ modes. 
The complete description of pairing vibration excitations in heavy-ion collisions might also be possible with such a code. 

\section{Deep-inelastic collisions \label{sec:DIC}}

Deep-inelastic collisions (DIC) have been widely studied in the past \cite{roy77,sch77,sch84,fre84,eva91}. 
They are characterised by a strong damping of the relative kinetic energy and an orbiting of the di-nuclear system before re-separation of the two fragments. 
In particular, large widths of the fragment mass and charge distributions are usually observed.

DIC have been used to investigate isospin equilibration in damped collisions \cite{pla88,pla90,sou88,sou89}, and to produce nuclei and study their structure (see for example Refs.~\cite{for98,lan01}). 
Upcoming radioactive beams will be used to further investigate transport properties of
isospin asymmetric nuclear matter \cite{lem12}. 

The characteristics of DIC provide stringent tests to nuclear transport models \cite{dav78,ran82,wei80}.
For instance, it has been shown that standard TDHF calculations usually fail to reproduce the large widths of mass and charge distributions \cite{koo77,dav78}. 
This is due to the fact that the many-body wave-function is constrained to be a single Slater determinant at all time \cite{das79}.
Fluctuations are then computed with beyond TDHF approaches such as the TDRPA \cite{bal84} and stochastic mean-field \cite{ayi08} formalisms. 

We first discuss briefly the calculation of fluctuations at the TDRPA level. 
Then, we present applications to the $^{40}$Ca+$^{40}$Ca reaction well above the barrier.

\subsection{Fluctuations of one-body observables}

Balian and V\'en\'eroni (BV) have introduced a variational principle in which the TDHF theory turns out to be optimised to the expectation value of one-body observables \cite{bal81}. 
It could then fail to reproduce quantities like two-body observables and fluctuations of one-body observables. 

Balian and V\'en\'eroni  also used their variational principle to derive a prescription for fluctuations and correlations between one-body observables \cite{bal84,bal92} (a detailed derivation can also be found in Ref.~\cite{sim12b}). 
This prescription is, in fact, fully equivalent to the TDRPA approach where small fluctuations around the mean-field evolution are considered. 

The BV variational principle can then be used 
to determine an optimum prediction for correlations $\sigma_{XY}$ and fluctuations $\sigma_{XX}$ 
of one-body operators assuming small fluctuations around the mean-field path \cite{bal84,bal92}.
Correlations are obtained from the general expression
\oeq
\sigma_{XY}^2=\lll\frac{1}{2}\(\<\oX\oY\>+\<\oY\oX\>\)-\<\oX\>\<\oY\>\rll,
\label{eq:corr_def}
\ceq
where $\sigma_{XY}$ has the sign of the term between the absolute value bars.
The $X$ and $Y$ distributions are correlated (resp. anti-correlated) for $\sigma_{XY}>0$ (resp. $\sigma_{XY}<0$). 
Fluctuations are obtained by taking $\oX=\oY$, leading to
\oeq
\sigma_{XX}=\sqrt{\<\oX^2\>-\<\oX\>^2}.
\ceq

Assuming independent particle states, the BV variational principle leads to the prescription
\oeq
\sigma_{XY}^2(t_1)=\lim_{\epsilon\rightarrow0}\frac{1}{2\epsilon^2}\tr \{\[\ro(t_0)-\ro_X(t_0,\epsilon)\]\[\ro(t_0)-\ro_Y(t_0,\epsilon)\]\} ,
\label{eq:corr}
\ceq
where $\tr$ denotes a trace in the single-particle space. 
The one-body density matrices $\ro_X(t,\epsilon)$ obey 
the TDHF equation~(\ref{eq:TDHF}) with the boundary condition at the final time~$t_1$
\oeq
\ro_X(t_1,\epsilon)=e^{i\epsilon q_X}\ro(t_1)e^{-i\epsilon q_X},
\ceq
while $\ro(t)=\rho_X(t,0)$ is the standard TDHF solution. 

Eq.~(\ref{eq:corr}) has been solved numerically in the past for particle number fluctuations with simple effective interactions and geometry restrictions~\cite{mar85,bon85,tro85}.
Modern three-dimensional TDHF codes with full Skyrme functionals are now used for realistic applications of the BV variational principle to determine these fluctuations~\cite{bro08,bro09,sim11,sim11b,sim12b} as well as the correlations between the proton and neutron number distributions \cite{sim11,sim11b}. 
See Ref.~\cite{sim12b} for numerical details of the implementation of Eq.~(\ref{eq:corr}). 

\subsection{The $^{40}$Ca+$^{40}$Ca reaction well above the barrier}

\begin{figure}
\includegraphics[width=11.5cm]{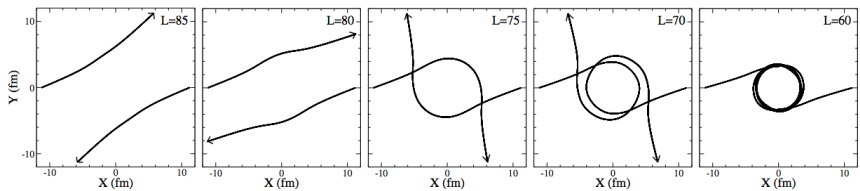}
\caption{Trajectories of the fragment centers-of-mass in $^{40}$Ca+$^{40}$Ca at $E_{c.m.}=128$~MeV and different angular momenta~$L$ in units of~$\hb$. 
\label{fig:CaCatraj}}
\end{figure}

The $^{40}$Ca+$^{40}$Ca reaction has been investigated at $E_{c.m.}=128$~MeV ($\sim2.4$ times the barrier)  \cite{sim11}  with the \textsc{tdhf3d} code and its TDRPA extension using the SLy4$d$ Skyrme functional \cite{kim97}. 
Fig.~\ref{fig:CaCatraj} provides some examples of trajectories obtained at different angular momenta. 
The TDHF calculations for this reaction predict that fusion occurs at $L\le66~\hb$ \cite{sim11}. 
We see in Fig.~\ref{fig:CaCatraj} that orbiting followed by re-separation is predicted at $L=70\hb$. 
Partial orbiting  at $L=75\hb$  and smaller nuclear deflections at larger $L$ are also observed. 

It is interesting to note that different angular momenta may lead to similar scattering angles.
This is the case, for instance, with  $L=70\hb$ and $L=75\hb$ which are associated to different orbiting trajectories (see Fig.~\ref{fig:CaCatraj}). 
In fact, the amount of orbiting changes very rapidly with $L$ for DIC. 
Comparisons with experimental data imply then to perform calculations with a small angular momentum step in the orbiting region \cite{sim11}. 

\begin{figure}
\sidecaption
\includegraphics[width=7.5cm]{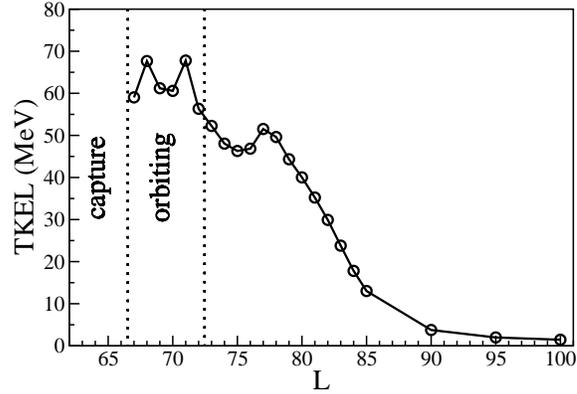}
\caption{Total kinetic energy loss of the fragments in $^{40}$Ca+$^{40}$Ca at $E_{c.m.}=128$~MeV as a function of the  angular momentum~$L$ in units of~$\hb$.   Adapted from Ref.~\cite{sim11}.  
\label{fig:CaCaTKE}}
\end{figure}

As mentioned before, DIC are not only characterised by a large orbiting of the di-nuclear system.
They are also associated to a large damping of the initial relative kinetic energy. 
This is quantified by the total kinetic energy loss $TKEL=E_{c.m.}-E_1-E_2$, where $E_{1,2}$ are the  asymptotic kinetic energies of the fragments in the exit channel. 
The $TKEL$ in $^{40}$Ca+$^{40}$Ca at $E_{c.m.}=128$~MeV are shown in Fig.~\ref{fig:CaCaTKE} as  a function of the initial angular momentum.
The maximum $TKEL$ of $\sim60-70$~MeV are obtained close to the critical angular momentum for fusion. 
These $TKEL$ have to be compared with the Viola systematics for fission fragments \cite{vio85}. 
The latter gives an expected $TKEL$ in symmetric fission of $\sim$76~MeV.
This indicates that the DIC around $L\simeq70\hb$ are almost fully damped. 
Note that this result is obtained with TDHF calculations which contain one-body dissipation only. 
As a result, the damping of relative kinetic energy in DIC is essentially of one-body nature.

Another characteristic of DIC is the large width of the fragment mass and charge distributions. 
Independent particle descriptions such as the  TDHF theory usually strongly underestimate these widths \cite{koo77,dav78,sim11}.
It is then necessary to include beyond TDHF fluctuations with the TDRPA \cite{bal84} or with the stochastic mean-field (SMF) approach \cite{ayi08}.
Calculations with the TDRPA \cite{mar85,bon85,tro85,bro08,bro09,sim11,sim11b} and with the SMF approach \cite{was09b,yil11} indeed predict larger fluctuations than the TDHF theory. 

\begin{figure}
\sidecaption
\includegraphics[width=7.5cm]{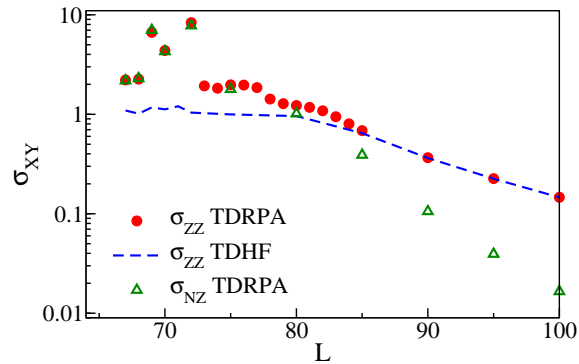}
\caption{Fragment charge fluctuation $\sigma_{ZZ}$ with TDHF (dashed line) and TDRPA (filled circles) as a function of the angular momentum $L$ in units of~$\hb$. The fluctuations $\sigma_{NN}$ of the neutron number distributions, not shown, are very close to the proton ones. Correlations $\sigma_{NZ}$ between proton and neutron numbers distributions obtained with TDRPA calculations are shown with open triangles. Adapted from Ref.~\cite{sim11}.  
\label{fig:CaCasigma}}
\end{figure}

This is illustrated in Fig.~\ref{fig:CaCasigma} where the fragment charge fluctuations obtained from TDHF (dashed line) and TDRPA (filled circles) are reported as a function of the angular momentum for the $^{40}$Ca+$^{40}$Ca reaction at $E_{c.m.}=128$~MeV \cite{sim11}.
We see that, for large angular momenta $L\ge90\hb$, both approaches predict similar fluctuations. 
These reactions are very peripheral and associated to small $TKEL$ of few MeV (see Fig.~\ref{fig:CaCaTKE}). 
Collisions at $L\ge90\hb$ are then dominated by quasi-elastic scattering. 
This shows that the TDHF approach may be used safely to compute quasi-elastic transfer (see also section~\ref{sec:transfer}) as it provides similar fluctuations than the TDRPA for these quasi-elastic events.

\vspace{3cm}

\begin{figure}
\sidecaption
\includegraphics[width=7.5cm]{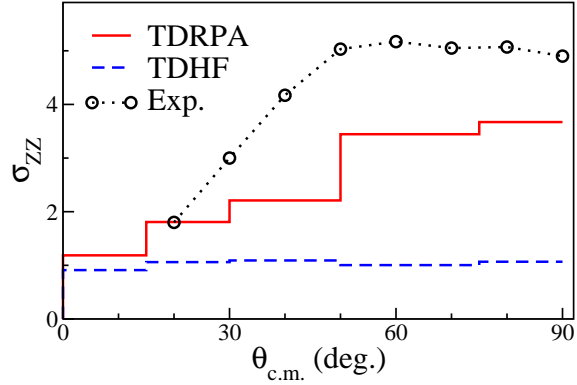}
\caption{Fluctuations $\sigma_{ZZ}$ of the proton number distributions obtained with TDHF (dashed line) and TDRPA (solid line) calculations are plotted as function of the center-of-mass scattering angle $\theta_{c.m.}$.   Experimental data (open circles) from Ref.~\cite{roy77} are also indicated. Adapted from Ref.~\cite{sim11}.  
\label{fig:CaCasig_tet}}
\end{figure}

On the other hand, we observe in Fig.~\ref{fig:CaCasigma} that fluctuations with the TDRPA are much larger than with the TDHF approach for more central collisions, in particular in the DIC region at $L\sim70\hb$.
A comparison of these fluctuations with the experimental data of Roynette and collaborators \cite{roy77} has been performed in Ref.~\cite{sim11}. 
The results are reported in Fig.~\ref{fig:CaCasig_tet} for a selection of  events with $TKEL>30$~MeV. 
Although the TDRPA results still underestimate experimental data, they provide both a better qualitative and quantitative agreement than the TDHF calculations. 
In fact, the plateau observed at large angles contains a contribution from fusion-fission events \cite{roy77}. 
The latter are not treated in the calculations and may be the origin of the remaining difference between the TDRPA calculations and the experimental data~\cite{sim11}.

Correlations between proton and neutron numbers distributions have also been computed recently with the TDRPA approach for  $^{40}$Ca+$^{40}$Ca collisions at $E_{c.m.}=128$~MeV \cite{sim11}. 
In standard TDHF calculations, these correlations are strictly zero. 
In the TDRPA, however, they become important in the DIC region, as shown in Fig.~\ref{fig:CaCasigma} (open triangles). 
Although they are negligible for quasi-elastic scattering ($L\ge90\hb$), they are similar to the charge fluctuations for the most damped events. 
This indicates that, e.g., an addition of several protons in one fragment is likely to be accompanied by an addition of neutrons as well. 
This is a manifestation of the symmetry energy which favours $N=Z$ fragments.

Finally, combining neutron and proton fluctuations with their correlations, one can estimate the distribution of nuclei produced in the reaction. 
Let us assume a Gaussian probability distribution of the form
\oeq
\mathcal{P}(n,z) = \mP(0,0)\exp\[ -\frac{1}{1-\rho^2} \( \frac{n^2}{\sigma_N^2}+\frac{z^2}{\sigma_Z^2} - \frac{2\rho nz}{\sigma_N\sigma_Z}\) \], 
\ceq
where $n$ and $z$ are the number of transferred neutrons and protons, respectively.  
The probability for the inelastic channels reads
\oeq
\mP(0,0) = \(2\pi\sigma_N\sigma_Z\sqrt{1-\rho^2}\)^{-1}
\nonumber 
\ceq
The dimensionless quantity
\oeq
\rho = \mbox{sign}(\sigma_{NZ})\frac{\sigma_{NZ}^2}{\sigma_N\sigma_Z}=\frac{\<nz\>}{\sqrt{\<n^2\>\<z^2\>}}
\ceq
quantifies the correlations and obeys $|\rho|<1$. 
The case $\rho=0$ corresponds to independent distributions of the form $\mP(n,z)=\mP(n)\mP(z)$. 
On the other side, fully (anti-)correlated distributions are found in the limit $\rho\rightarrow\pm1$.

\begin{figure}
\sidecaption
\includegraphics[width=7.5cm]{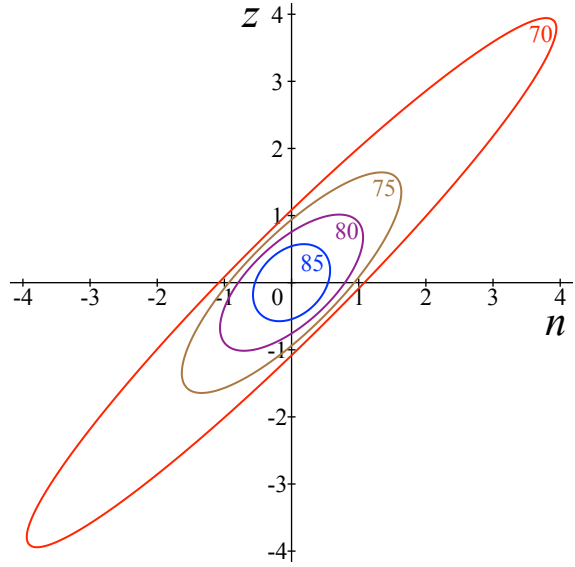}
\caption{TDRPA calculations of iso-probabilities $\mP(n,z)=\mP(0,0)/2$ for $n$ neutrons and $z$ protons transfered in $^{40}$Ca+$^{40}$Ca at $E_{c.m.}=128$~MeV. The numbers on the curves indicate the angular momentum in units of~$\hb$. \label{fig:CaCacorrel}}
\end{figure}

Iso-probabilities corresponding to $\mP(n,z)=\mP(0,0)/2$ are plotted in Fig.~\ref{fig:CaCacorrel} for different angular momenta. 
Independent proton and neutron distributions would produce ellipses with the principal axis parallel to the abscissa or to the ordinate. 
This is not what is observed, particularly for the smallest angular momenta corresponding to the most violent collisions. 
We see that not only the fluctuations are important to determine  distributions of DIC, but the correlations play a significant role as well.

To conclude, the TDHF theory is a good tool to compute transfer probabilities in quasi-elastic scattering. 
However, beyond TDHF fluctuations are mandatory to describe fragment mass and charge distributions in more violent reactions such as deep-inelastic collisions. 
Calculations based on the TDRPA indeed provide a better agreement with experimental data than standard TDHF codes. 
The correlations are also  shown to be important in DIC with these TDRPA calculations. 
They should be sensitive to the symmetry energy  and might be used in the future to test this part of the functional, in particular with exotic beams. 

\section{The quasi-fission process \label{sec:QF}}

When two nuclei collide with an energy above the Coulomb barrier and a small enough impact parameter, a capture of the two fragments is expected to occur, i.e., a di-nuclear system is formed after dissipation of the relative kinetic energy \cite{ada12}. 
The outcome of such a di-nuclear system is either $(i)$ fusion, i.e., the formation of a unique system where two-centers cannot be identified anymore in the density distribution, or $(ii)$ a re-separation after a possible multi-nucleon transfer between the fragments. 

In light and medium mass systems, fusion is usually enabled by a close contact between the fragments. 
For heavy systems with typical charge products  greater than $\sim$1600-1800, however, the second process is often dominant around the Coulomb barrier, leading to a fusion hindrance in these systems \cite{gag84}. 
Instead of fusing, the di-nuclear system encounters a re-separation in two fragments after a possible exchange of a large number of nucleons. 
This process is called quasi-fission as the characteristics of the fragments may exhibit some strong similarities with those emitted in statistical fission of the compound nucleus formed by fusion \cite{boc82,tok85,she87}. 
Note that, although much less probable than fusion, quasi-fission may also occur in lighter systems \cite {hin96,itk11,nas07}. 

Firstly, we present some TDHF calculations of fusion hindrance in several heavy systems. 
Then, we investigate the effect of some structure properties of the collision partners, in particular their deformation, on the quasi-fission process. 

\subsection{Fusion hindrance in heavy systems}

\begin{figure}
\sidecaption
\includegraphics[width=3cm]{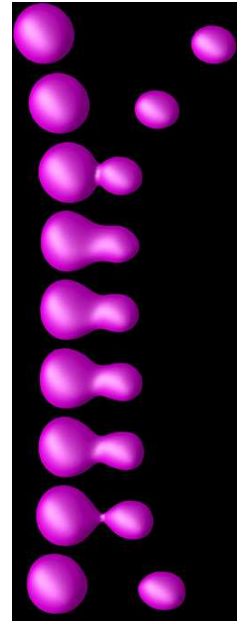}
\caption{Snapshots of the density for a  $^{56}$Fe+$^{208}$Pb central collision at $E_{c.m.}=240$~MeV. 
The isodensity is plotted at half the saturation density, i.e., at $\rho_0/2=0.08$~fm$^{-3}$.
Snapshots are shown every 0.75~zs. 
Time runs from top to bottom.  
\label{fig:FePbdens}}
\end{figure}

Let us illustrate the phenomenon of fusion hindrance with TDHF calculations of heavy systems using the \textsc{tdhf3d} code with the SLy4$d$ Skyrme functional \cite{kim97}. 
Fig.~\ref{fig:FePbdens} shows the density evolution of a $^{56}$Fe+$^{208}$Pb ($Z_1Z_2=2132$) central collision at $E_{c.m.}=240$~MeV.
This energy is well above the Coulomb barrier.
Indeed, the barrier computed with the proximity model \cite{blo77} is $B_{prox.}\simeq224$~MeV. 
However, this collision does not lead to fusion.
Indeed, despite the formation of a rather compact di-nuclear system, the latter end up in quasi-fission. 
The lifetime of the di-nuclear system is $\sim3$~zs, which is too short to enable a full mass equilibration~\cite{boc82,tok85,she87,rie11}  which would be signed by a symmetric exit channel.

Fusion may eventually occur in some collisions if enough energy above the Coulomb barrier is brought into the system. 
This additional energy, sometimes called ''extra-push'' energy, may be computed with phenomenological approaches such as the extra-push model of Swiatecki \cite{swi82}. 
Modern TDHF calculations are also able to determine such fusion thresholds without any parameter adjusted on reaction mechanism \cite{sim12b,sim12c,guo12}. 

\vspace{0.5cm}

\begin{figure}
\sidecaption
\includegraphics[width=7.5cm]{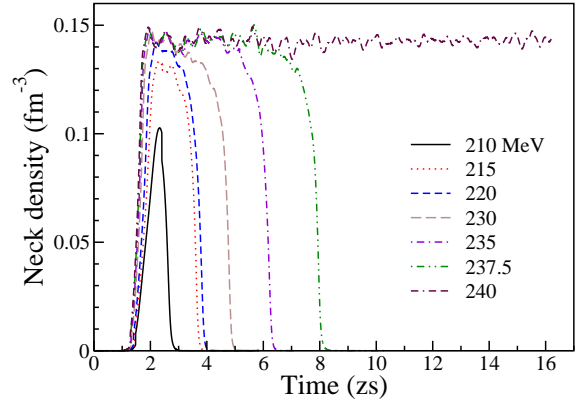}
\caption{Neck density in  $^{90}$Zr+$^{124}$Sn central collisions as function of time \cite{ave09c}. The center-of-mass energies are indicated in the legend. 
\label{fig:neck}}
\end{figure}

An example is provided in Fig.~\ref{fig:neck} where the neck density is plotted as a function of time for $^{90}$Zr+$^{124}$Sn ($Z_1Z_2=2000$) central collisions at several energies. 
This system has a proximity barrier $B_{prox.}\simeq215$~MeV.
Densities exceeding $0.13$~fm$^{-3}$ are observed at this energy and above. 
However, re-separation occurs for energies smaller than 240~MeV.
Increasing contact times with energy are observed below 240~MeV. 
At $E_{c.m.}=240$~MeV, however, the neck survives more than 14~zs, which may be interpreted as a fusion process. 
An extra-push energy of 22 to 25~MeV above the proximity barrier is then needed for the $^{90}$Zr+$^{124}$Sn system to fuse. 

\begin{figure}
\sidecaption
\includegraphics[width=7.5cm]{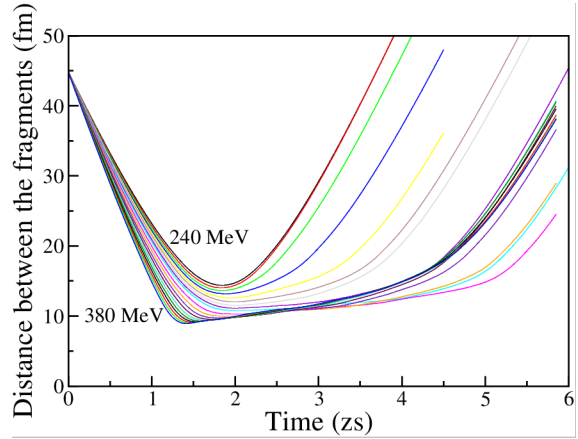}
\caption{Distance between the centers-of-mass of the fragment in $^{70}$Zn+$^{208}$Pb central collisions as a function of time for center-of-mass energies between 240~MeV and 380~MeV \cite{ave09c}.  
\label{fig:ZnPb}}
\end{figure}

\begin{figure}
\sidecaption
\includegraphics[width=7.5cm]{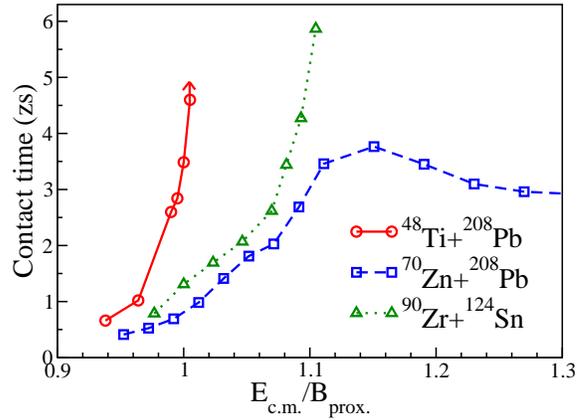}
\caption{Contact time between the fragments arbitrarily defined as the time the systems spend at a distance between their centers-of-mass smaller than 15~fm. Adapted from Ref.~\cite{sim12d}. The arrow indicates a lower limit.  
\label{fig:contact_heavy}}
\end{figure}

Increasing the collision energy does not guarantee to reach such a fusion threshold in all systems.
For instance, TDHF calculations predict that fusion is not possible in the $^{70}$Zn+$^{208}$Pb ($Z_1Z_2=2460$) system \cite{sim12d}.
This is illustrated in Fig.~\ref{fig:ZnPb} where the distance between the centers-of-mass of the fragments is plotted as a function of time at different center-of-mass energies ranging from 240~MeV to 380~MeV.
The proximity barrier for this system is $B_{prox.}\simeq252$~MeV. 
We see that, at an energy of more than 100~MeV above this barrier, the system is still not fusing. 
In fact, we observe a rise and fall of the contact time in this system with increasing energy in Fig.~\ref{fig:contact_heavy}. 
This indicates that no fusion is expected in this system. 

\vspace{1cm}

To conclude, the fusion hindrance in heavy systems which hinders the formation of very heavy elements by fusion is described in the TDHF approach.
When not fusing, the di-nuclear systems encounter quasi-fission within several zeptoseconds. 
Fusion may occur in some systems with an additional extra-push energy. 
However, others, with larger charge products are never found to fuse whatever the energy. 

\subsection{Effects of the structure of the collision partners}

The previous section shows the importance of quasi-fission in the outcome of heavy di-nuclear systems. 
Realistic descriptions of quasi-fission are challenging because many degrees of freedom are at play. 
In addition, the shape of the di-nuclear system and its mass and isospin repartition evolves dynamically on different time scales. 

The equilibration of the isospin degree of freedom has been studied with the TDHF approach in several works \cite{bon81,sim01,sim07,uma07,iwa09,iwa10a,iwa10b,iwa12,obe12,sim11,sim12a}. 
It has been shown to occur on a typical time scale smaller than 2~zs \cite{sim12a}. 
This time scale is smaller than standard quasi-fission times. 
Quasi-fission fragments are then expected to have a similar $N/Z$ ratio than the compound nucleus. 

Shell effects of the collision partners have also been shown to play an important role in the competition between fusion and quasi-fission \cite{sto98,faz05,sim12a}. 
In particular, fusion might be eased by the magicity as less dissipation is expected in the fusion valley of magic nuclei \cite{hin05,arm00}. 
As a result, more compact di-nuclear systems can be formed \cite{san76,gup77,faz05,ari06}. 
This may explain the success of super-heavy element synthesis with the doubly-magic $^{48}$Ca projectile \cite{oga06,hof07}. 
However, these effects remain to be investigated with quantum microscopic approaches such as the TDHF formalism.

The role of deformation and orientation on quasi-fission has been investigated in several experiments \cite{hin95,liu95,hin96,oga04,kny07,hin08,nis08}. 
These studies led to the general conclusion that quasi-fission in collisions with the tip of a deformed heavy target (e.g., in the actinide region) is dominant. 
On the other side, collisions with the side lead to more compact shapes which favour long lifetimes of the di-nuclear systems and then increase the fusion probability. 

The effect of deformation and orientation on quasi-fission has been investigated recently with TDHF calculations \cite{sim12b,sim12c,sim12d,wak12} of the $^{40}$Ca+$^{238}$U system.
In particular, it has been shown that collisions with the tip of $^{238}$U do not lead to fusion, but to quasi-fission with a contact time of $\sim10$~zs  and a partial mass equilibration almost independent with energy. 
On the contrary, contact with the side of $^{238}$U  leads to long di-nuclear lifetimes above the barrier which may induce the formation of a compound nucleus. 

In Ref.~\cite{sim12a}, it was shown that $^{40}$Ca and $^{48}$Ca behave differently as far as the interplay between quasi-fission and fusion is concerned when colliding with a $^{208}$Pb target. 
In particular, the hindrance of quasi-fission due to shell effects is observed only with $^{48}$Ca, despite the fact that $^{40}$Ca is also doubly magic. 
This is interpreted as an effect of isospin asymmetry which, in the case of $^{40}$Ca, induces a fast $N/Z$ equilibration breaking the magicity of the fragments in the di-nuclear system \cite{sim12a}.

\vspace{1cm}

\begin{figure}
\sidecaption
\includegraphics[width=7.5cm]{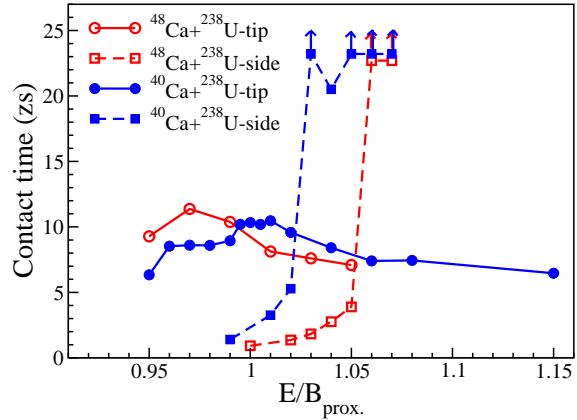}
\caption{Contact time between the fragments as function of center-of-mass energy normalised by the proximity barrier \cite{blo77} in $^{40,48}$Ca+$^{238}$U central collisions. Here, contact times are defined as the time during which the neck density exceeds half the saturation density $\rho_0/2=0.08$~fm$^{-3}$. Collisions with the tip (solid line) and the side (dashed line) of the $^{238}$U are considered. The arrows indicate lower limits.  
\label{fig:time_CaU}}
\end{figure}

It is  interesting to see if  differences in the quasi-fission process between the two calcium isotopes are also observed in collision with a heavy deformed nucleus. 
Contact times, defined, here, as the time during which the neck density exceeds $\rho_0/2=0.08$~fm$^{-3}$, have been computed  in $^{48}$Ca+$^{238}$U central collisions with the \textsc{tdhf3d} code and the SLy4$d$ Skyrme functional \cite{kim97}. 
The evolution of these contact times as a function of energy is plotted in Fig.~\ref{fig:time_CaU} for collisions with the tip (solid line) and with the side (dashed line) of $^{238}$U. 
The behaviours of the contact times present similarities between the two isotopes. 
For instance, quasi-fission times for collisions with the tip are of the order of 10~zs for both $^{40}$Ca and $^{48}$Ca. 
Collisions with the side also present a sharp increase of the contact time above the barrier in both cases, with long contact times (more than 20~zs) at high energy which could lead to fusion. 
However, long contact times for collisions with the side are reached at higher energies with $^{48}$Ca than with $^{40}$Ca. 
This might be attributed to  shell effects in the $^{48}$Ca-like fragment which are absent in the $^{40}$Ca-like fragment due to $N/Z$ equilibration \cite{sim12a}. 
This effect needs further investigations. 

To sum up, the competition between quasi-fission and fusion is affected by the structure of the collision partners. 
In particular, the deformation and the orientation is crucial. 
Recent TDHF calculations confirm that collisions with the tip lead essentially to quasi-fission, while long contact times possibly leading to fusion may be reached above the barrier for collisions with the side producing to more compact shapes. 
More investigations with quantum microscopic approaches are needed to gain a deeper understanding on how the various structure characteristics,  such as, e.g., shell effects and isospin, affect quasi-fission. 

\section{Actinide collisions \label{sec:actinides}}

Collisions of actinides form ''hyper-heavy molecules'' with $\sim500$ nucleons in interaction during short times of few zeptoseconds \cite{zag10}.
The description of their dynamics is of course a great challenge for theorists. 
These reactions may be an alternative way to produce more neutron-rich heavy and super-heavy  nuclei than those formed by fusion \cite{vol78,zag06,ked10,sim11b}. 
This is possible thanks to the fact that actinides have large neutron to proton ratio, of the order of $N/Z\sim1.5$. 

Another interest of actinide collisions is the possibility to make the QED vacuum unstable due to the strong electric field \cite{rei81,gre83,ack08}.
As a result, a spontaneous decay of the vacuum by the emission of a $e^+e^-$ pair is expected. 
The lifetime of the hyper-heavy molecule is a crucial parameter which  determines if this QED vacuum decay may be observed experimentally. 

These  applications of actinide collisions require a precise description of the reaction mechanisms. 
The dynamics of actinide collisions has been investigated with various theoretical approaches, including macroscopic models \cite{zag06,fen09,sar09}, semi-classical microscopic approaches \cite{tia08,zha09}, and the TDHF theory \cite{cus80,str83,gol09,ked10}. 

In the following we describe, first, the role of the relative orientation of the nuclei on the reaction mechanisms. 
Then, we look for the optimal conditions for the observation of spontaneous $e^+e^-$ emission. 

\subsection{Role of the initial orientation}

Di-nuclear systems formed in actinide collisions are too heavy to fuse.
They always encounter quasi-fission.
As in quasi-fission with lighter projectiles, the deformation and orientation of the nuclei play a crucial role in the outcome of the collision \cite{zag10}. 
This has been confirmed with recent TDHF calculations of the $^{238}$U+$^{238}$U \cite{gol09} and $^{232}$Th+$^{250}$Cf \cite{ked10} systems with the \textsc{tdhf3d} code and the SLy4$d$ Skyrme functional~\cite{kim97}. 

\begin{figure}
\begin{center}
\includegraphics[width=11.5cm]{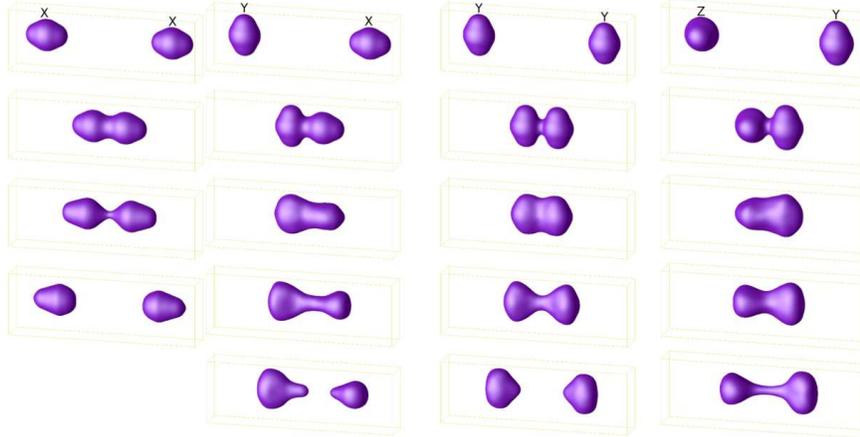}
\end{center}
\caption{Isodensities at half the saturation density, i.e., $\rho_0/2=0.08$~fm$^{-3}$, in $^{238}$U+$^{238}$U central collisions at a center-of-mass energy $E_{c.m.}=1200$~MeV. Evolutions associated to the four initial configurations $XX$, $YX$, $YY$, and $ZY$ are plotted in columns (time runs from top to bottom). Consecutive snapshots are separated by 1.125~zs. Adapted from Ref.~\cite{sim11b}
\label{fig:U+Udensity}}
\end{figure}

As an example, Fig.~\ref{fig:U+Udensity} shows snapshots of the density in $^{238}$U+$^{238}$U central collisions at $E_{c.m.}=1200$~MeV. 
Different shape evolutions are observed depending on the initial orientations of the actinides. 
In particular, a collision of the tips ($XX$ configuration) leads to a rapid neck formation, but to a faster re-separation of the fragments than with the other orientations. 
The most compact configurations are obtained for side on side collisions ($YY$ and $ZY$ configurations). 
In particular, the $ZY$ configuration leads to the longest contact times as it has less Coulomb repulsion than the $YY$ orientation. 

These orientations exhibit also differences as far as the mass flow between the di-nuclear fragments is concerned. 
In fact, no net transfer is observed due to symmetry reasons, except when a tip collide with a side ($YX$ configuration).
Indeed, in this case, no spatial symmetry prevents a net mass transfer to occur between the fragments. 
In fact, TDHF calculations predict that a large amount of nucleons are transferred from the tip to the side, allowing for the production of neutron-rich fragments in the fermium ($Z=100$) region in the $^{238}$U+$^{238}$U reaction at energies around the Coulomb barrier \cite{gol09}.
This phenomenon has also been investigated with the $^{232}$Th+$^{250}$Cf system using the TDHF approach. 
In particular, it is shown that when the tip of the $^{232}$Th collide with a side of the $^{250}$Cf, the latter increases its mass, producing new neutron-rich transfermium nuclei \cite{ked10,sim11b}. 
This phenomenon is called ''inverse quasi-fission'' as the exit channel is more mass asymmetric than the colliding partners. 
Note that inverse quasi-fission is also expected from shell effects in the $^{208}$Pb region, as shown by calculations based on the Langevin equation \cite{zag06} (See chapter~7 of {\it Clusters in nuclei Vol. 1} \cite{zag10}).

\begin{figure}
\sidecaption
\includegraphics[width=7.5cm]{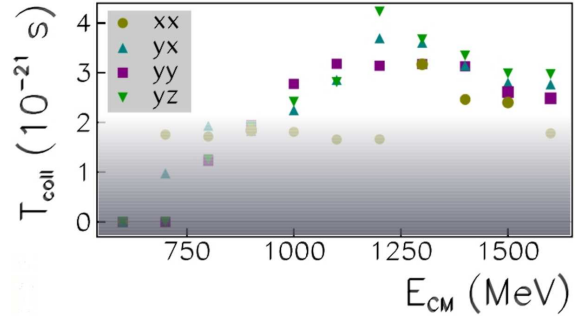}
\caption{Collision times for each orientation as a function of the center-of-mass energy. Here, these times are defined as the time during which the neck density exceeds $\rho_0/10=0.016$~fm$^{-3}$. The shaded area indicates the limit of 2~zs above which vacuum decay is expected to be observable in central collisions. Adapted from Ref.~\cite{gol09}. 
\label{fig:U+Utime}}
\end{figure}

\begin{figure}
\sidecaption
\includegraphics[width=7.5cm]{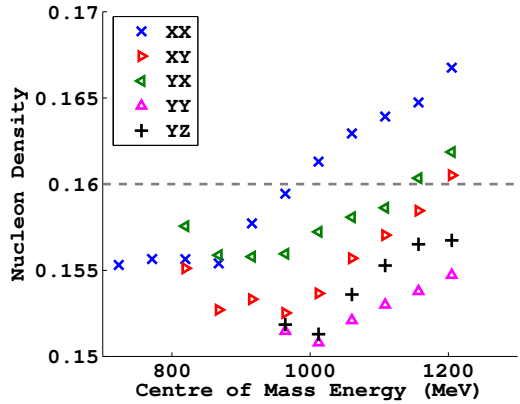}
\caption{Maximal density in the neck as a function of the center-of-mass energy in $^{232}$Th+$^{250}$Cf central collisions with different orientations (see text). The dashed line represents the saturation density at $\rho_0=0.16$~fm$^{-3}$. Adapted from Ref.~\cite{ked10}
\label{fig:Th+Cf_neck}}
\end{figure}

Let us now investigate in more details the role of the orientation on collision times.
Fig.~\ref{fig:U+Utime} gives the evolution of the contact time between the fragments in $^{238}$U+$^{238}$U central collisions as a function of energy and for the different orientations represented in the top of Fig.~\ref{fig:U+Udensity}. 
A saturation of the contact time to 2~zs is observed for tip on tip collisions ($XX$) up to $E_{c.m.}\simeq1200$~MeV. 
(Above this energy, $XX$ contact times increase due to ternary quasi-fission \cite{gol09}.) 
This saturation is interpreted as a repulsive force generated by large densities in the neck when the tips overlap.
A similar effect is observed in $^{232}$Th+$^{250}$Cf \cite{ked10}. 
The maximal density in the neck region for this system is reported in Fig.~\ref{fig:Th+Cf_neck}.
We observe that the $XX$ configuration leads  to large densities above the saturation density. 

\begin{figure}
\begin{center}
\includegraphics[width=5cm]{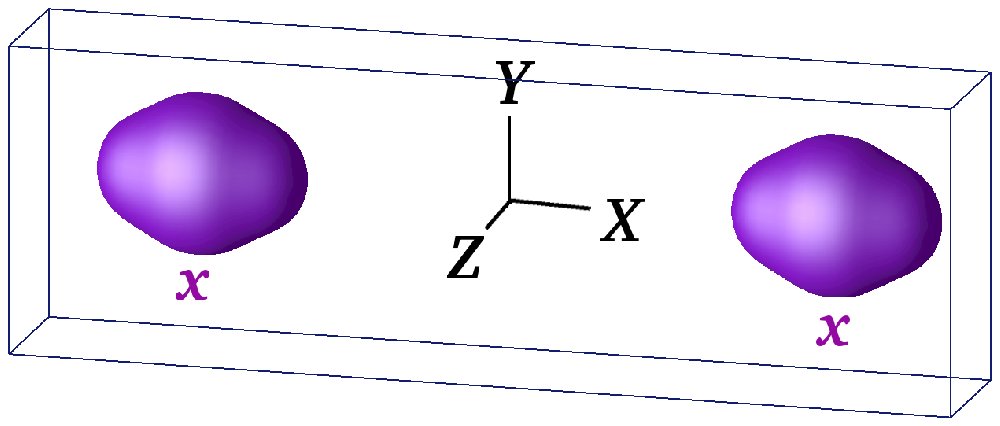}\\
\includegraphics[width=5cm]{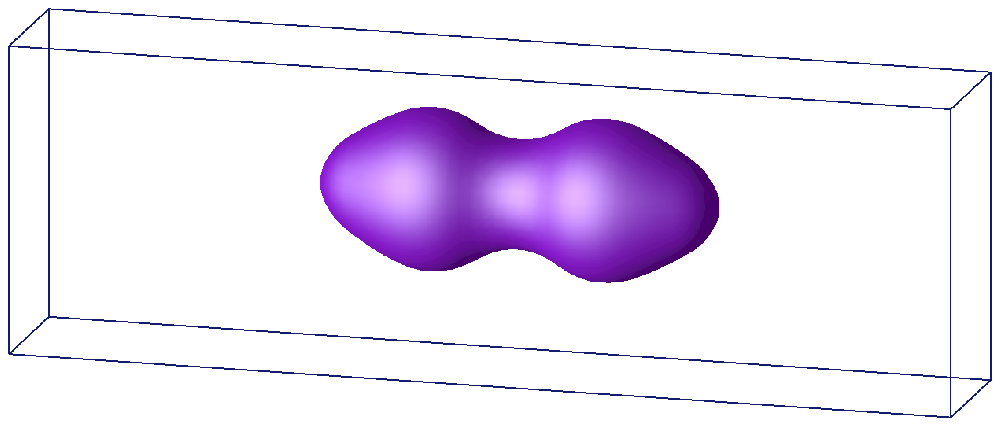}\\
\includegraphics[width=5cm]{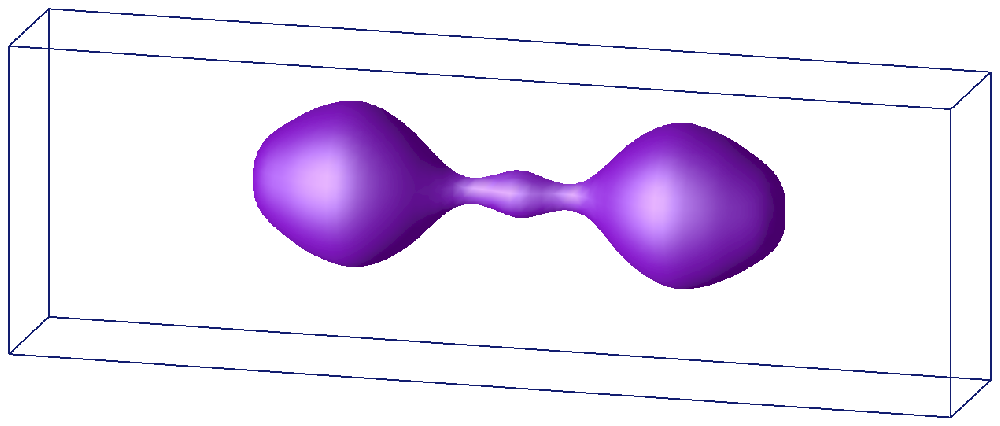}\\
\includegraphics[width=5cm]{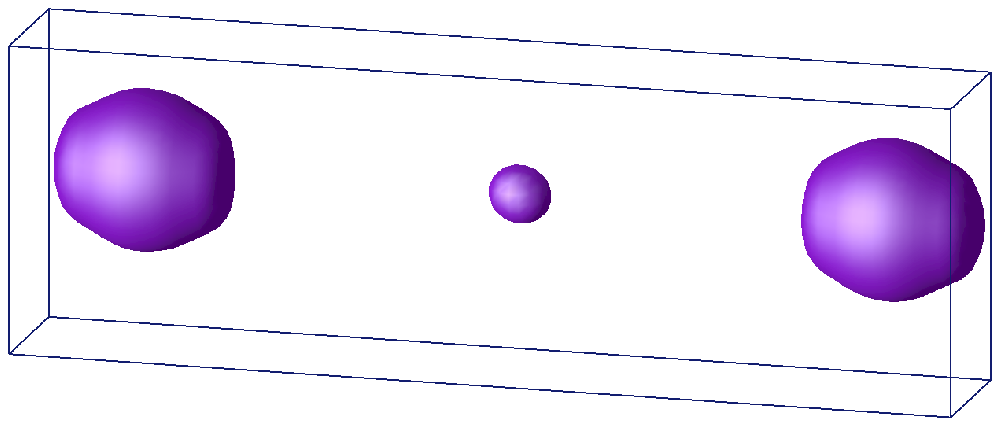}
\end{center}
\caption{Zeptosecond snapshots of $^{238}$U+$^{238}$U at $E_{c.m.}=900$~MeV for a tip on tip central collision. Adapted from Ref.~\cite{gol09}. 
\label{fig:U+Uternary}}
\end{figure}

\begin{figure}
\sidecaption
\includegraphics[width=7.5cm]{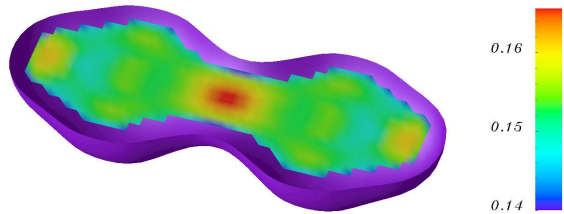}
\caption{Nucleon density (in fm$^{-3}$) in the collision plane is plotted when the density in the neck reaches its maximum in the $XX$ configuration at $E_{c.m.}=900$~MeV. The half cut surface is an isodensity at
half the saturation density, i.e. $\rho_0/2=0.08$~fm$^{-3}$. Adapted from Ref.~\cite{gol09}. 
\label{fig:U+Uneck}}
\end{figure}

These large densities in the neck are also responsible, in some cases, for a re-separation of the system in three fragments \cite{gol09}. 
This is illustrated in Fig.~\ref{fig:U+Uternary} for the $^{238}$U+$^{238}$U central tip on tip collision at $E_{c.m.}=900$~MeV. 
A small fragment is observed at rest in the exit channel.
A similar phenomenon is discussed in Ref.~\cite{zag10}. 
The formation of this third fragment is interpreted as due to an excess of density in the neck region. 
To illustrate this argument, Fig.~\ref{fig:U+Uneck} shows the internal density at the distance of closest approach. 
Densities above the saturation density are indeed observed in the neck region. 
In fact, instead of breaking in the middle of the neck, the system breaks in both sides of this over saturation density region, producing a third small fragment at rest \cite{gol09}.  

To conclude, the dynamics of actinide collisions has been studied with the TDHF approach. 
These systems encounter quasi-fission with typical di-nuclear system lifetimes of 2 to 4~zs, depending on the initial orientation of the nuclei. 
The mass transfer also strongly depends on the relative orientation at contact. 
In particular, collisions where the tip of a nucleus is in contact with the side of its collision partner lead to important mass transfer. 
These multi-nucleon transfer reactions may be used in the future for the production of neutron-rich transfermium nuclei. 
Finally, it is shown that the complex neck dynamics may lead to the production of a third fragment at rest. 

\subsection{Lifetime and spontaneous $e^+e^-$ emission}

\begin{figure}
\begin{center}
\includegraphics[width=6cm]{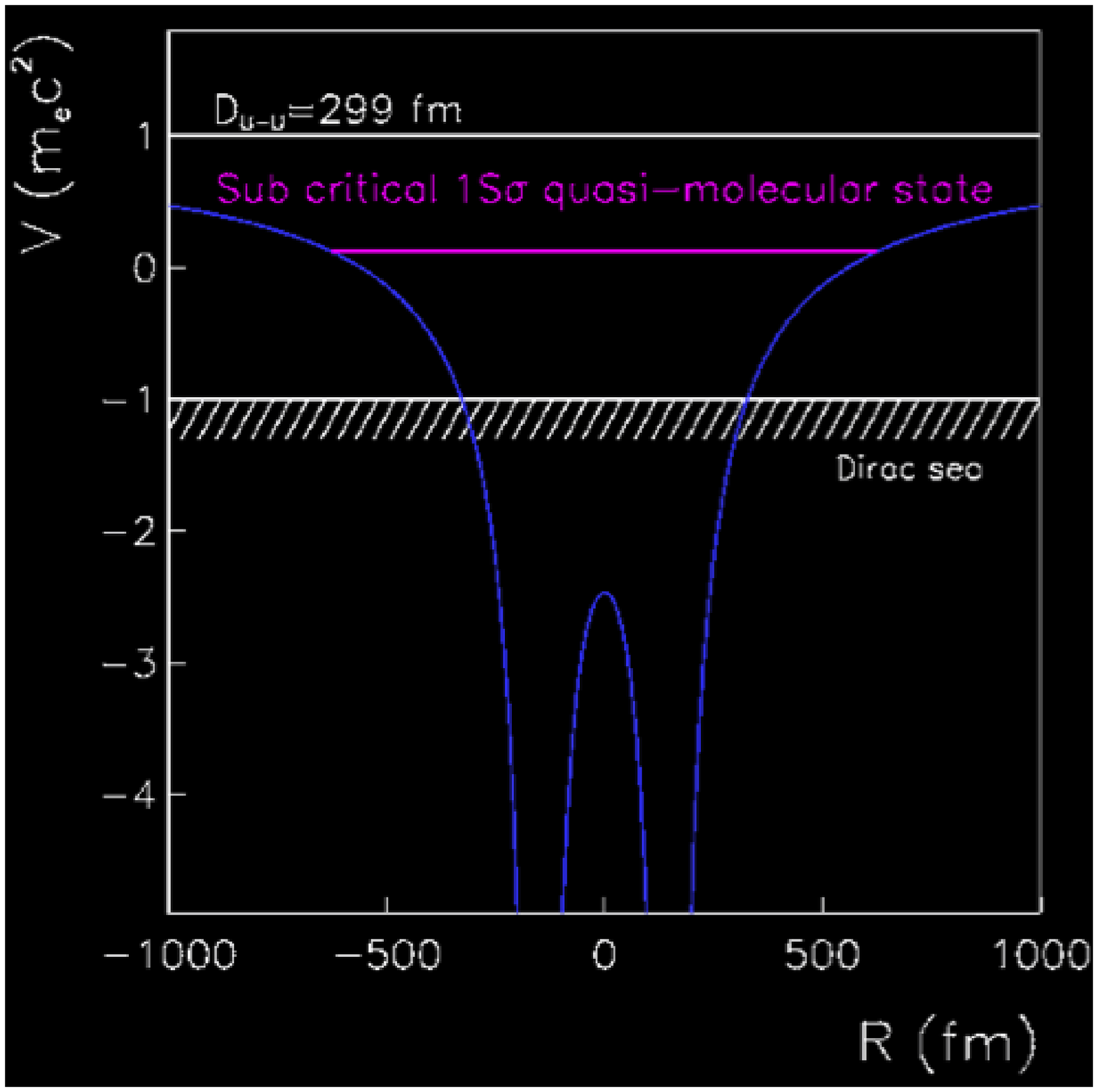}\\
\includegraphics[width=6cm]{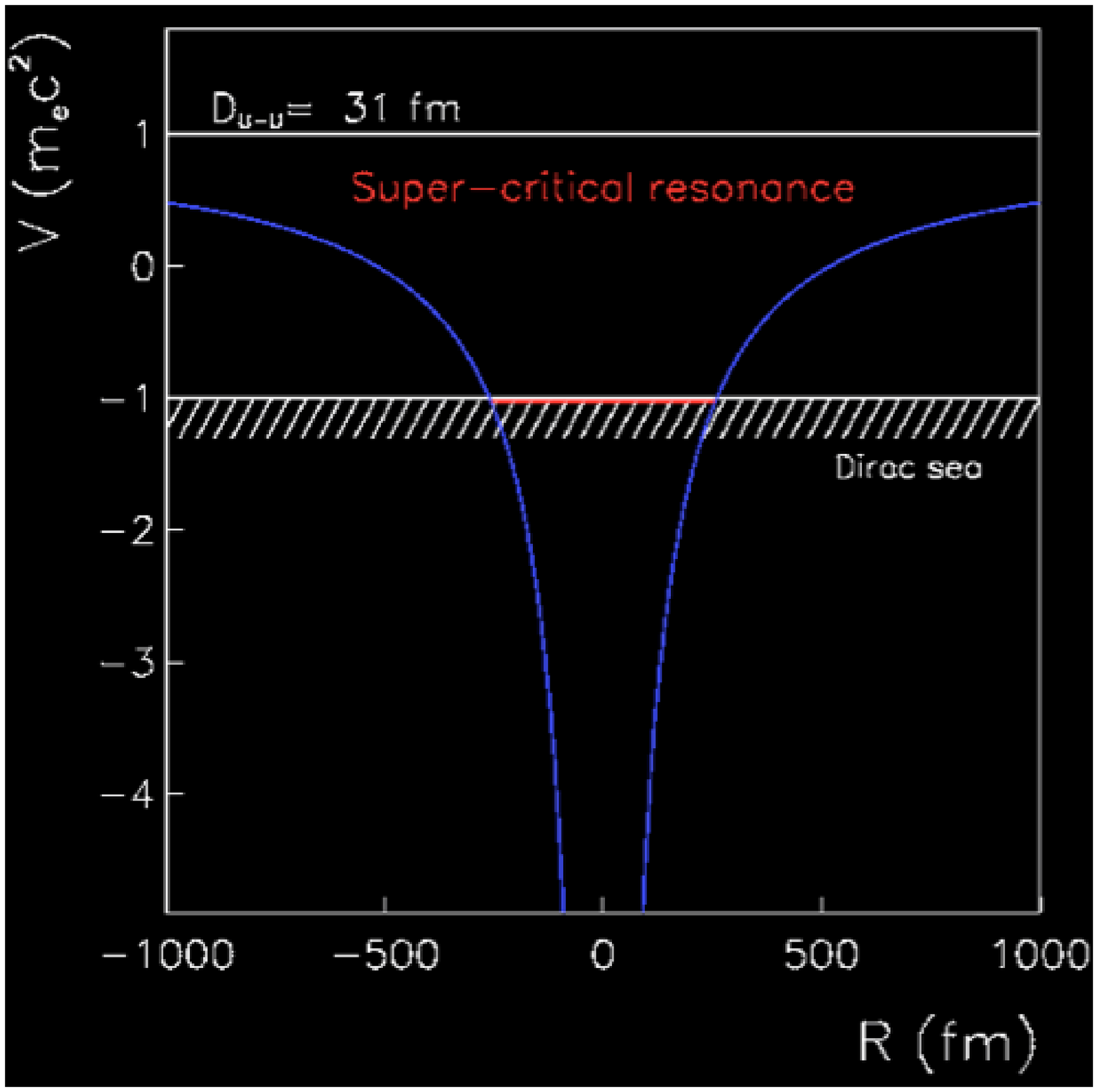}\\
\includegraphics[width=6cm]{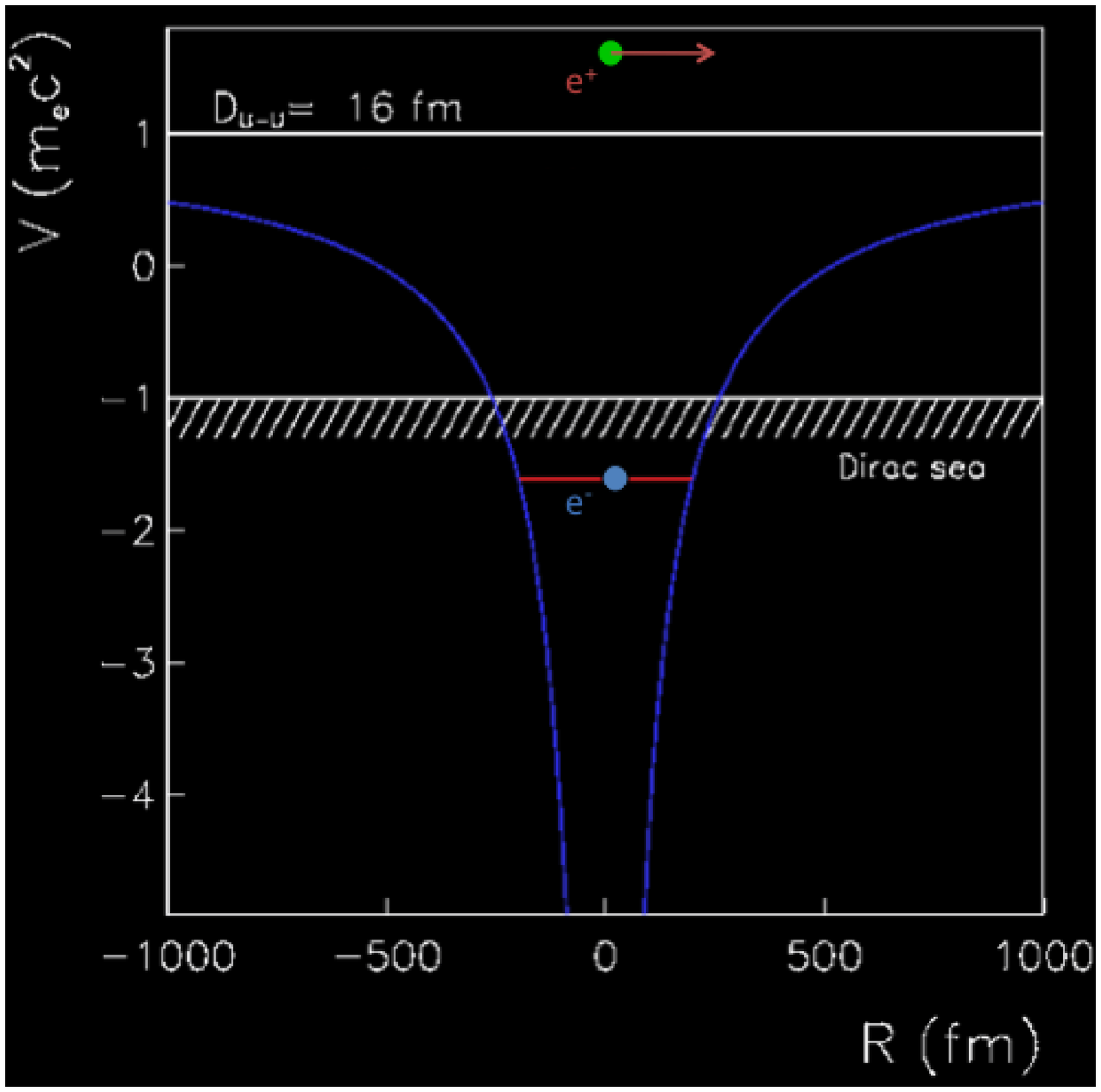}
\end{center}
\caption{Coulomb potential energy (blue solid lines) generated by two fully stripped uranium nuclei at distances $D_{U-U}=299$, 31 and 16~fm. The horizontal purple and red lines indicate the lowest quasi-molecular electronic state \cite{rei81}.  
\label{fig:QED}}
\end{figure}

It is predicted that nuclear systems with more than $\sim173$ protons generate super-critical Coulomb fields \cite{zag10,rei81,gre83,ack08}, i.e., with the lowest quasi-molecular electronic state in the Dirac sea. 
This is illustrated in Fig.~\ref{fig:QED} in the case of a collision of two bare uranium. 
Without Coulomb field, the lowest energy of an electron is $E=m_ec^2$ for an electron at rest. 
In the top of Fig.~\ref{fig:QED}, we see that when the two uranium are at a distance $D_{U-U}\simeq300$~fm, the lowest quasi-molecular state available for an electron has an energy $E\simeq0$ \cite{rei81}. 
At this distance, the system is still sub-critical as the state is above the Dirac sea. 

A super-critical state is obtained at $D_{U-U}\simeq36$~fm when the quasi-molecular state crosses the Dirac sea at $E=-m_ec^2$ (middle panel in Fig.~\ref{fig:QED}).
At this distance the nuclei are not yet in contact. 
The latter occurs at $D_{U-U}\simeq16$~fm.
At this distance, the quasi-molecular state has an energy $E\simeq-1.5m_ec^2$.

If the super-critical state is not or partially occupied, then it induces a hole in the Dirac sea. 
According to QED, such a state is unstable, i.e., it is a resonance with a finite lifetime depending on the depth of the energy level. 
This resonance is predicted to decay by producing a $e^+e^-$ pair (see bottom panel in Fig.~\ref{fig:QED}). 
In this case, the $e^-$ occupies the  state with negative energy $E$, while the positron is emitted in the continuum with an energy $-E$.

\begin{figure}
\sidecaption
\includegraphics[width=7.5cm]{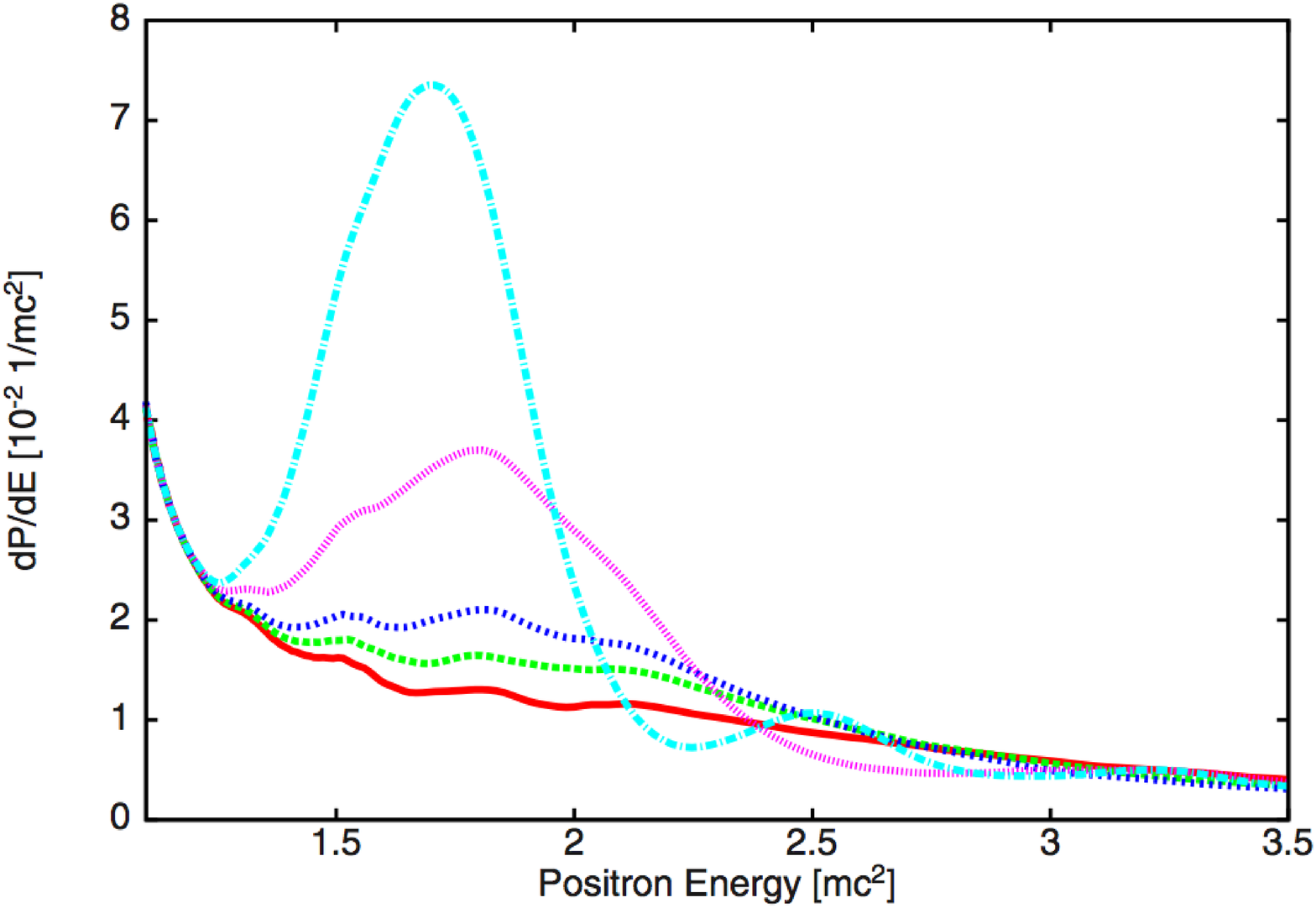}
\caption{Total differential positron spectrum $dP/dE$ for the U-U system at $E_{c.m.}=740$~MeV, for different nuclear sticking times calculated using the time-dependent Dirac equation. The solid red curve is for $T=0$~zs, the long-dashed green curve for $T=1$~zs, the short-dashed blue curve for $T=2$~zs, the dotted magenta curve for $T=5$~zs, and the dash-dotted cyan curve for $T=10$~zs. Adapted from Ref.~\cite{ack08}. 
\label{fig:positron}}
\end{figure}

The Dirac equation has been solved to predict the lifetime of the resonance in Ref.~\cite{ack08}.
Fig.~\ref{fig:positron} shows the resulting positron spectra for different hypotheses about the contact time $T$ between the actinides. 
The positrons emitted in the case $T=0$ (solid red line) are due to standard $e^+e^-$ pair creation when electric charges are accelerated in a strong Coulomb field. 
For finite contact times, additional positrons are produced from the decay of the vacuum. 
These additional positrons form a peak around $E=1.5-2mc^2$ which can be observed for contact times greater than 2~zs. 

To enable the observation of the QED vacuum decay via $e^+e^-$ pair emission, it is then crucial to have  contact times between the actinides as large as possible. 
Predictive calculations of the nuclear dynamics, in particular of the contact times between the nuclei are then mandatory.  
We see in Fig.~\ref{fig:U+Utime} that contact times greater than 2~zs are obtained for central collisions at $E_{c.m.}\ge1000$~MeV. 
In particular, the largest contact times of 3-4~zs  obtained at $E_{c.m.}\simeq1200$~MeV might enable the observation of spontaneous QED vacuum decay. 
Similar conclusions have been reached with  quantum molecular dynamics calculations \cite{tia08}. 
Note that these energies are greater than the previous experimental search for spontaneous $e^+e^-$ emissions.
Indeed, these experiments were using beams around the Coulomb barrier at 700-800~MeV (see Ref.~\cite{ahm99} and references therein). 

To sum up, a possible signature of the QED vacuum decay would be the observation of a peak in the  spectra of positrons produced in actinide collisions. 
Microscopic calculations predict optimum center-of-mass energies at more than 1~GeV.
These energies are much larger than the energies considered in early experiments. 

\section{Dynamics of neutron star crust \label{sec:crust}}

Although the description of supernova explosion mechanisms is not yet complete \cite{col66}, it is well known that ''neutron stars'' are a possible residue of such explosions.
These dense baryonic objects have drawn lots of interests in the past \cite{hae07}. 
However, their structure, which is intimately linked to the nuclear phase diagram, is still actively discussed nowadays.
For instance, the order of the crust-core phase transition is  under debate \cite{rad10,gul11}.
In addition, the structure of the inner core may be affected by possible phase transitions involving strangeness \cite{gul12}. 

\begin{figure}
\begin{center}
\includegraphics[width=12cm]{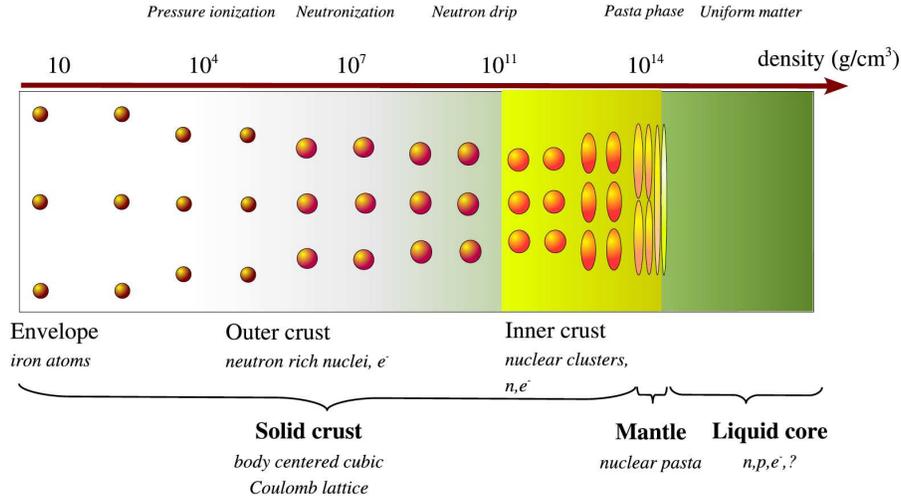}
\end{center}
\caption{Schematic picture of the structure of a neutron star as a function of density. From Ref.~\cite{cha08} \label{fig:schema}}
\end{figure}

The outer layers of neutron stars are also expected to exhibit exotic structures involving different geometrical organisations of the nuclear species \cite{oya93}.
Fig.~\ref{fig:schema} provides a schematic picture of neutron stars \cite{cha08}.
The outer crust  is expected to exhibit a lattice of droplets in a sea of nuclear matter \cite{bay71,rav83,mas84}. 
In this picture, the inner crust is composed of nuclear clusters which may be unstable to quadrupole deformations, forming prolate nuclei.
The latter would eventually join up to form stringlike structures  in the so-called ''nuclear pasta phase'' \cite{pet95}. 

The total energy is shared between bulk, surface, and Coulomb contributions. 
Inside nuclei, the density is approximatively constant.
As a result, only the surface and Coulomb terms depend on nuclear shapes.
The details of the diagram in Fig.~\ref{fig:schema} are then sensitive to a subtle competition between the surface tension, which makes nuclei spherical, versus the Coulomb force, which tends to deform them. 

\begin{figure}
\sidecaption
\includegraphics[width=7.5cm]{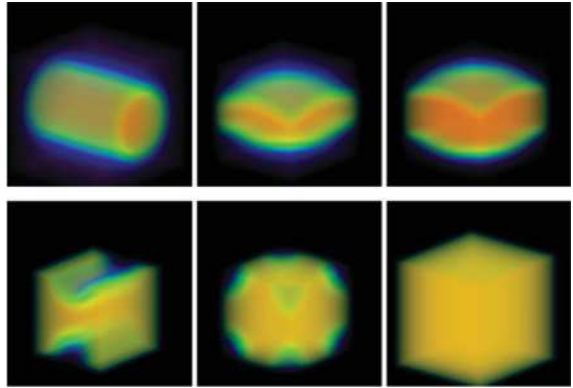}
\caption{Neutron density profiles at a temperature $T=2.5$~MeV, a proton fraction $x_p=0.3$, and baryon densities of $0.04$~fm$^{-3}$ (top left), 0.06~fm$^{-3}$ (top middle), 0.08~fm$^{-3}$ (top right), 0.09~fm$^{-3}$ (bottom left), 0.10~fm$^{-3}$ (bottom middle), and 0.11~fm$^{-3}$  (bottom right). The lowest neutron densities are in dark (blue) colors, while the highest are in gray (red). From Ref.~\cite{new09} \label{fig:crust_HF}}
\end{figure}

Static Hartree-Fock calculations based on Skyrme functionals have been performed to investigate these structures in more details in Refs.~\cite{mag02,gog07,new09}. 
As an example, Fig.~\ref{fig:crust_HF} represents neutron densities for different total baryon densities from 0.04 to 0.11~fm$^{-3}$ computed by Newton and Stone \cite{new09}.
Except at the highest density, where an homogeneous distribution is obtained, various pasta phases can be observed. 

The formation of these structures has been also investigated within microscopic theories \cite{wat02,wat09,seb09,seb11}. 
For instance, the quantum molecular dynamics (QMD) semi-classical model has been used by Watanabe and collaborators to show that pasta phase could be formed dynamically \cite{wat02,wat09}. 

\begin{figure}
\sidecaption
\includegraphics[width=7.5cm]{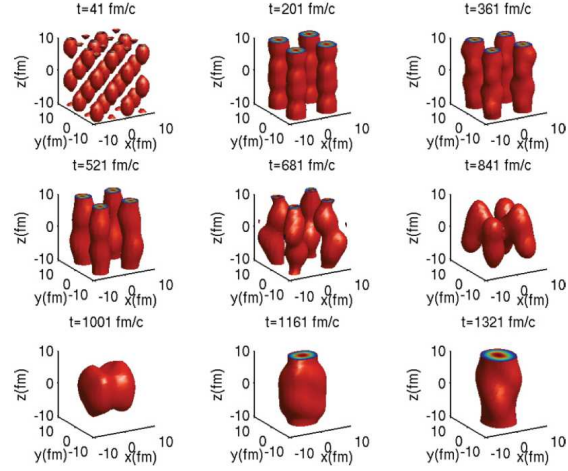}
\caption{Snapshots of the density profiles of a perturbed supercell of oxygen isotopes in a face-centered cubic lattice. The mean neutron density is $\<\rho\>=0.058$~fm$^{-3}$, and  a proton fraction $x_p=0.5$ is considered. From Ref.~\cite{seb11} 
\label{fig:FCC}}
\end{figure}

In addition, using a quantum microscopic framework based on the TDHF equation (called the DYWAN model), S\'ebille and collaborators have investigated the stability of some lattice structures \cite{seb09,seb11}. 
As an example, the evolution of an initially perturbed face-centered oxygen lattice with proton fraction $x_p=0.5$ and mean neutron density $\<\rho\>=0.058$~fm$^{-3}$ toward a cylinder is shown in Fig.~\ref{fig:FCC}. 
The initial perturbation consists of a random displacement of the oxygen centers in order to break the mean-field symmetry. 

\begin{figure}
\sidecaption
\includegraphics[width=7.5cm]{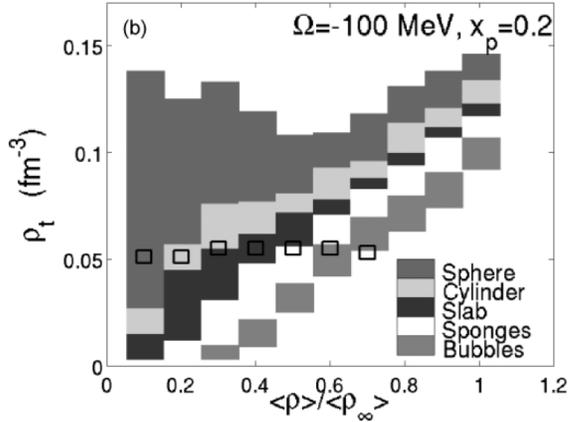}
\caption{Neutron structure distributions as a function of the threshold density $\rho_t$ at which isodensities are plotted and of the neutron mean density $\<\rho\>$ normalised to $\rho_\infty=0.145$~fm$^{-3}$ for a proton fraction $x_p=0.2$. Initial simple cubic lattices of oxygen isotopes are used. 
 From Ref.~\cite{seb11} 
\label{fig:phases}}
\end{figure}

The role of proton fraction, lattice geometry, and nuclear species on the lattice sites have also been considered in Ref.~\cite{seb11}. 
Different shapes of the pasta phase have then been obtained, depending on the mean neutron density $\<\rho\>$ and on the threshold density $\rho_t$  used to represent the isodensities. 
An example of distributions of structures as a function of these quantities is shown in Fig.~\ref{fig:phases}.
We see that the TDHF mean-field dynamics is able to reproduce the different types of structures expected in pasta phases.

To conclude, the dynamics of nuclear cluster aggregation and pasta phase formation from regular lattice of nuclei can be studied at the TDHF level. 
It should be noted, however, that these dynamics may be sensitive to the energy-density functional, and, in particular, to its isospin and density dependences \cite{mar07}. 
In addition, the neutron structure distributions could be different if non-local effective forces were used \cite{chappert08,loa11}.
Finally, the role of beyond mean-field correlations remains to be investigated.

\section{Selected conclusions and perspectives}

The TDHF approach provides a  mean-field description of nuclear dynamics in the presence of some clustering and molecular effects. 
The same formalism is used to investigate light systems exhibiting $\al$-clustering, heavy-ion collisions in a wide energy and mass range, and instabilities of lattice of nuclei in neutron star crusts. 

Non-trivial dynamics are observed depending on the initial conditions. 
For instance, $\al$-clusters, which are shown to survive and vibrate several zeptoseconds in light systems such as $^4$He+$^8$Be$\rightarrow^{12}$C, quickly dissolve when entering the mean-field of a heavy nucleus.  
Another example is the formation of di-nuclear systems in collisions of atomic nuclei. 
These systems are possible doorways to the formation of molecular states of the compound system. 

The dynamics of di-nuclear systems is crucial as it determines the outcome of the reaction, i.e., a re-separation of the fragments or their fusion in a compound nucleus. 
Although they exhibit some common features across the nuclear chart, such as lifetimes of the order of few zeptoseconds, the evolution of di-nuclear systems may strongly depend on the entrance channel properties. 
For instance, fusion probabilities are much larger in light systems, while the heaviest ones formed in actinide collisions always encounter quasi-fission.
It is also shown that deformation and orientation are playing an important role in the dynamics. 

In terms of theoretical description of the proton-neutron composition of the final fragments, the TDHF approach can be safely used for the less violent collisions, such as quasi-elastic transfer reactions, although one should keep in mind that only sequential transfer is included, i.e., transfer of clusters of nucleons is neglected.
However, damped reactions, such as deep-inelastic collisions and quasi-fission, usually involve a large transfer of nucleons between the fragments which may be under-estimated at the mean-field level. 
In fact, experimental charge and mass distributions in damped collisions are usually wider than TDHF predictions. 
This drawback is overcame thanks to the inclusion of fluctuations at the TDRPA level. 
In addition, not only particle number fluctuations are important in damped collisions, but also correlations between proton and neutron distributions are shown to be large. 
These correlations, which can be computed within the TDRPA approach, are crucial for a good description of fragment distributions in damped collisions. 

The transfer of correlated nucleons (paired nucleons, $\al$-clusters...) is one of the main challenges for microscopic approaches. 
For instance, the inclusion of pairing correlations at the TDHFB level will help to describe pair transfer reactions.  
In particular, a possible Josephson effect similar to what is observed in superconductors could be observed. 

A good understanding of radiative capture between light nuclei at deep sub-barrier energies is crucial to describe stellar nucleosynthesis. 
However, the present time-dependent mean-field approaches do not enable a tunneling of the many-body wave-function. 
Beyond TDHF approaches should then be considered to describe sub-barrier fusions of astrophysical interest. 
In particular, dynamical long-range fluctuations must be included. 
Possible candidates are the time-dependent GCM, where the generator coordinate is the distance between the fragments \cite{des12}, and the imaginary time-dependent Hartree-Fock formalism derived from the Feynman path-integral approach for many-body systems \cite{neg82}. 

\begin{acknowledgement}

Long term theoretical collaborations with D. Lacroix, Ph. Chomaz, and B. Avez are acknowledged.
Some of the calculations have been performed in collaboration with B. Avez, C. Golabek, D.~J. Kedziora, D. Lebhertz, and A. Wakhle.
The redaction of this chapter would not have been possible without numerous scientific exchanges with  theoreticians and experimentalists. 
In particular, discussions with E. Ackad, G. Adamian, N. Antonenko, Ch. Beck, M. Bender, K. Bennaceur, S. Courtin, M. Dasgupta, T. Duguet, M. Evers, W. Greiner, F. Gulminelli, F. Haas, D.~J. Hinde, D. Jacquet, D.~H. Luong, J. Meyer, J.~A. Maruhn, M. Morjean, V. de la Mota, M.-G. Porquet, R. du Rietz, N. Rowley, M.-D. Salsac, P. Schuck, F. S\'ebille, E. Suraud, and S. Umar are acknowledged. 
E. Ackad, N. Chamel, M. Evers, J.~A. Maruhn, V. de la Mota, W. Newton, F. S\'ebille, and S. Umar are warmly thanked for providing figures of their own works. 
The calculations with the \textsc{tdhf3d} code were performed on the Centre de Calcul Recherche et Technologie of the Commissariat \`a l'\'Energie Atomique, France, and on the NCI National Facility in Canberra, Australia, which is supported by the Australian Commonwealth Government.
Some of the works presented in this chapter were performed with the support from ARC grants DP0879679, DP110102858, DP110102879, and FT120100760.

\end{acknowledgement}

\bibliographystyle{epj}
\bibliography{biblio}

\end{document}